\documentclass[compsoc, conference, a4paper, 10pt, times]{IEEEtran}
\usepackage{cite}
\usepackage{amsmath,amssymb,amsfonts}
\usepackage{graphicx}
\usepackage{textcomp}
\usepackage{booktabs}
\usepackage[hidelinks]{hyperref}
\usepackage[table]{xcolor}
\usepackage{capt-of}

\usepackage{amsfonts}
\usepackage{subfig}
\usepackage{wrapfig}
\usepackage{float}
\usepackage{multirow, graphicx}
\usepackage{algorithm}
\usepackage{algpseudocode} 
\usepackage{amsmath}
\usepackage{siunitx}
\usepackage{textcomp}

\usepackage{xspace} 

\usepackage{url}

\usepackage{stackrel}

\ifdefined\TTSTYLETARGETreport
\usepackage[
    backend=bibtex,
    maxbibnames=100,
    firstinits=true,
    bibstyle=numeric,
    citestyle=numeric
]{biblatex}
\fi

\usepackage{pifont}
\newcommand{\cmark}{\ding{51}}%
\newcommand{\xmark}{\ding{55}}%

\usepackage{enumitem}

\usepackage{todonotes}
\usepackage{graphicx, multirow}

\usepackage{tabularx, multirow, rotating} 
\usepackage{subfig} 
\usepackage{algorithm} 
\usepackage{algpseudocode} 
\usepackage{ulem} 
\usepackage{color} 

\usepackage{listings} 
\newcommand{\lstuppercase}{\uppercase\expandafter{\expandafter\lst@token
                           \expandafter{\the\lst@token}}}
\newcommand{\lstlowercase}{\lowercase\expandafter{\expandafter\lst@token
                           \expandafter{\the\lst@token}}}
\makeatother

\usepackage{tikz}

\newif\ifboldnumber

\algrenewcommand\alglinenumber[1]{%
  \footnotesize\ifboldnumber\bfseries\fi\global\boldnumberfalse#1:}

\begin{document}

\title{Understanding the Security Risks of Decentralized Exchanges in the Wild}
\title{
Understanding the Security Risks of Decentralized Exchanges by Uncovering Unfair Trades in the Wild 
}

\makeatletter
\newcommand{\linebreakand}{%
  \end{@IEEEauthorhalign}
  \hfill\mbox{}\par
  \mbox{}\hfill\begin{@IEEEauthorhalign}
}
\makeatother

\author{\IEEEauthorblockN{Jiaqi Chen}
\IEEEauthorblockA{\textit{Syracuse University} \\
jchen217@syr.edu}
\and
\IEEEauthorblockN{Yibo Wang}
\IEEEauthorblockA{\textit{Syracuse University} \\
ywang349@syr.edu}
\and
\IEEEauthorblockN{Yuxuan Zhou}
\IEEEauthorblockA{\textit{Syracuse University} \\
yzhou168@syr.edu}
\and
\IEEEauthorblockN{Wanning Ding}
\IEEEauthorblockA{\textit{Syracuse University} \\
wding04@syr.edu}
\linebreakand
\IEEEauthorblockN{Yuzhe Tang}
\IEEEauthorblockA{\textit{Syracuse University} \\
ytang100@syr.edu}
\and
\IEEEauthorblockN{XiaoFeng Wang}
\IEEEauthorblockA{\textit{Indiana University Bloomington} \\
xw7@indiana.edu}
\and
\IEEEauthorblockN{Kai Li\textsuperscript{\textsection}}
\IEEEauthorblockA{\textit{San Diego State University} \\
kli5@sdsu.edu}
}
\maketitle
\begingroup\renewcommand\thefootnote{\textsection}
\footnotetext{This work is done when Kai Li was a Ph.D. student at Syracuse University.}
\endgroup

\begin{abstract}
DEX, or decentralized exchange, is a prominent class of decentralized finance (DeFi) applications on blockchains, attracting a total locked value worth tens of billions of USD today. 

This paper presents the first large-scale empirical study that uncovers unfair trades on popular DEX services on Ethereum and Binance Smart Chain (BSC). By joining and analyzing $60$ million transactions, we find $671,400$ unfair trades on all six measured DEXes, including Uniswap, Balancer, and Curve. Out of these unfair trades, we attribute $55,000$ instances, with high confidence, to token thefts that cause a value loss of more than $3.88$ million USD. Furthermore, the measurement study uncovers previously unknown causes of extractable value and real-world adaptive strategies to these causes. Finally, we propose countermeasures to redesign secure DEX protocols and to harden deployed services against the discovered security risks.

\end{abstract}
\newcommand{\yibo}[1]{\footnote{\textcolor{red}{(Yibo's comment: #1)}}}
\newcommand{\kai}[1]{\footnote{\textcolor{red}{(Kai's comment: #1)}}}

\newcommand{\yuzhe}[1]{\footnote{\textcolor{red}{(Yuzhe: #1)}}}

\lstdefinestyle{mystyle}{
    backgroundcolor=\color{backcolour},   
    commentstyle=\color{codegreen},
    keywordstyle=\color{codepurple},
    numberstyle=\numberstyle,
    stringstyle=\color{blue},
    basicstyle=\footnotesize\ttfamily,
    breakatwhitespace=false,
    breaklines=true,
    captionpos=b,
    keepspaces=true,
    numbers=left,
    numbersep=10pt,
    showspaces=false,
    showstringspaces=false,
    showtabs=false,
}
\lstset{style=mystyle}

\newcommand\numberstyle[1]{%
    \footnotesize
    \color{codegray}%
    \ttfamily
    \ifnum#1<10 0\fi#1 |%
}

\definecolor{mygreen}{rgb}{0,0.6,0}
\lstset{ %
  backgroundcolor=\color{white},   
  basicstyle=\scriptsize\ttfamily,        
  breakatwhitespace=false,         
  breaklines=true,                 
  captionpos=b,                    
  commentstyle=\color{mygreen},    
  deletekeywords={...},            
  escapeinside={(*}{*)},          
  extendedchars=true,              
  keepspaces=false,                 
  keywordstyle=\color{blue},       
  language=Java,                 
  morekeywords={*,...},            
  numbers=left,                    
  numbersep=5pt,                   
  numberstyle=\scriptsize\color{black}, 
  rulecolor=\color{black},         
  showspaces=false,                
  showstringspaces=false,          
  showtabs=false,                  
  stepnumber=1,                    
  stringstyle=\color{purple},     
  tabsize=2,                       
  title=\lstname,                  
  moredelim=[is][\bf]{*}{*},
}

\definecolor{codegreen}{rgb}{0,0.6,0}
\definecolor{codegray}{rgb}{0.5,0.5,0.5}
\definecolor{codepurple}{HTML}{C42043}
\definecolor{backcolour}{HTML}{F2F2F2}
%
%
%

\newcommand{\ignore}[1]{}

\section{Introduction}

Recent years have witnessed the rise of Decentralized Finance (DeFi) on blockchains. A prominent class of DeFi applications is Decentralized Exchange (DEX) which allows traders to swap their assets from one type of token to another. 
On Ethereum, the most popular DEX services that account for more than 95\% of the market (incl. Uniswap~\cite{me:dex:uniswap,Adams2020UniswapVC}, Sushiswap~\cite{me:sushiswap:1,me:sushiswap:2}, Pancakeswap~\cite{me:pancakeswap:1}, Curve~\cite{me:curve}, and Balancer~\cite{me:balancer}) { all run AMM protocols or Automated Market Maker~\cite{me:etherscan:topdex}.} In an AMM, a trader account deposits one token to a smart-contract instance, called a pool, and withdraws a certain amount of other tokens from the pool. Besides token swaps, an AMM
pool supports the operation of adding/removing ``liquidity'' where liquidity-provider accounts deposit/withdraw tokens to/from the pool to increase/decrease token reserves in the pool and to support future trades\footnote{This paper uses terms ``token swap'' and ``trade'', interchangeably.}. 

AMM can be exploited to extract value illicitly. Well known AMM attacks include sandwich attacks and arbitrage~\cite{DBLP:conf/sp/DaianGKLZBBJ20,DBLP:journals/corr/abs-2101-05511,DBLP:conf/uss/TorresCS21,DBLP:journals/corr/abs-2105-02784,DBLP:conf/sp/ZhouQCLG21}. In both attacks, an adversarial trader sends crafted transactions and executes {\it multiple} trades against a victim pool. Illicit value can be extracted as the multiple trades can exploit the difference in token exchange rates at different times (i.e., in sandwich attacks~\cite{DBLP:journals/corr/abs-2101-05511}) or across different pools (i.e., in arbitrage~\cite{DBLP:journals/corr/abs-2105-02784}).


\begin{figure}[b!ht]
\begin{center}
\includegraphics[width=0.425\textwidth]{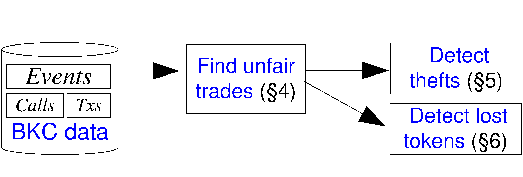}
\end{center}
\caption{Transaction-analysis workflow to discover unfair trades (described in \S~\ref{sec:detectviolatingswap}) and to attribute them into thefts (\S~\ref{sec:attack}) and lost tokens (\S~\ref{sec:losttokens}).}
\label{fig:workflow7}
\end{figure}

{%
\vspace{2pt}\noindent\textbf{Uncovering unfair trades}. 
This work aims at studying a different class of AMM exploits: Instead of abusing price differences across multiple trades, the attacker aims to incur and exploit the unfairness of a single trade.\footnote{This paper uses ``trade'' to represent all other operations an AMM may support, such as adding/removing liquidity.} 
Specifically, an AMM trade occurs when trader Alice wants to deposit token X to the AMM's pool and withdraw token Y from the pool. A trade is fair if the amount of token X Alice deposits and the amount of token Y she withdraws conforms to a trusted exchange rate. An unfair trade may occur in different forms, and unfairness could imply financial risks or even attacks. For instance, unfairness may manifest by a token swap being executed in non-atomic fashion, that is, a standalone withdrawal (i.e., without a matching deposit) or a standalone deposit (without a matching withdrawal). A standalone withdrawal could indicate token theft in which thief Alice is able to withdraw value from the victim pool without depositing. A standalone deposit could mean an adversarial AMM pool denying the legitimate withdrawal request from trader Alice who has made a deposit.
Besides non-atomicity, an unfair trade could manifest by other forms such as the deposit value and withdrawn value do not confirm to the exchange rate (value mismatch), and the pool receives deposit from trader Alice but allows another trader Bob to withdraw (account mismatch).\footnote{The fairness definition in this paper applies to a single trade, similar to existing works~\cite{DBLP:conf/podc/Herlihy18}. It should be differentiated from the multi-trade fairness as in MEV (\S~\ref{sec:rw}) and transaction-wise atomicity~\cite{DBLP:journals/corr/abs-2101-05511}. For instance, existing MEV attacks, such as sandwich attacks and arbitrage, do not violate the trade-wise fairness as in this work.}

By connecting together and analyzing over $60$ million transactions on Ethereum and Binance Smart Chain (BSC), to our surprise, we find unfair trades widely exist among deployed AMM services: 
Committing unfairness are the most popular services, including Uniswap V2~\cite{me:dex:uniswap,Adams2020UniswapVC}, Sushiswap~\cite{me:sushiswap:1,me:sushiswap:2}, Pancakeswap~\cite{me:pancakeswap:1}, Curve~\cite{me:curve}, and Balancer~\cite{me:balancer}.
In total, they have performed more than $671$ thousand AMM trades and liquidity operations violating fairness, exhibiting patterns including standalone token deposit, standalone withdrawal, account mismatch, and value mismatch.


\vspace{2pt}\noindent\textbf{Understanding unfair trades}. 
We aim to understand the nature of the unfair trades, whether they are accidents or intentional crimes. While unfair trades can be attributed to different causes (e.g., denial of service as mentioned above), we mainly focus on the prominent cases of token theft.

Detecting theft from unfair trades is challenging. This is because the ground truth often includes off-chain activities unaccounted for.
For instance, when attributing a standalone withdrawal to theft, one has to know that the party making the deposit to accounts and the party withdrawing from the accounts are controlled by two distinct physical entities, which is known to be an intractable problem with only on-chain data; known heuristics such as address clustering~\cite{DBLP:conf/fc/Victor20,DBLP:conf/socialcom/ReidH11,DBLP:conf/imc/MeiklejohnPJLMVS13,DBLP:conf/sp/HuangALIBMLLSM18,DBLP:books/daglib/0040621} do not apply here (see \S~\ref{sec:discoveryproblem} for detail).

To address the challenge, we propose three detection methods and cross-check them on each violation instance. First, we propose heuristics to match accounts' on-chain behavior against rational attack strategies. 
For instance, a rational thief, in fear of value extracted by competing thieves, would send crafted transactions with aggressively high fees and with short deposit-to-withdrawal delay, which can be used as a signature for detecting thefts.
Second, to eliminate the possibilities of benign value extraction, we propose detecting the frontrunning among multiple withdrawer accounts of the same pool, which shows the conflict of interest among withdrawals. Unlike existing works scanning the entire blockchain history and mempools~\cite{DBLP:conf/uss/TorresCS21,DBLP:conf/sp/DaianGKLZBBJ20}, we optimize the frontrunning detection by focusing on fairness-violating withdrawals and failed token withdrawals. Third, we also search the Internet for user complaints about our violation cases' accounts. The latter two methods jointly show ill-intention of the value extraction. 

Overall, our method is an end-to-end transaction-analysis pipeline, depicted in Figure~\ref{fig:workflow7}, which first discovers unfair trades and then detects attack instances in thefts and lost tokens.


\vspace{2pt}\noindent\textbf{Findings}. 
On the Ethereum mainnet and BSC fork, we discover $55$ thousand theft attacks mounted by $258$ distinct attacker accounts with high confidence, inflicting a loss of more than $3.88$ million USD stolen from $978$ victim accounts.
Each attack instance we found matches at least one indicator. 
As the measurement results show, real-world users exploit the undefined behavior in the AMM protocols, such as rebasing token supply, token interest and other non-standard token features, to successfully withdraw value. Interestingly, the attackers {\it adapt} their behavior to different exploits: When the attacker is to extract value from interest-bearing tokens, she waits long enough, typically hundreds of Ethereum blocks, for the interest to accumulate and to become profitable. Given other kinds of deposits, such as token rebase, the same attacker instead wastes no time extracting the illicit value, typically within one or two blocks, to prevent frontrunning.


\ignore{
We further study attackers' and victims' behavior by tracking their money flow and their actions to create/destroy smart contracts. Our study uncovers the systematic criminal activities: the crime bosses use mixer services to hide their identities while sending commissions and instructions. After the attack, the money is sent to various DeFi services to maximize illicit profit. Exploit smart contracts are widely used to improve attack success rates.
}

From the violations that cannot be attributed to theft, we further detect the instances of denial of service. Particularly, we formulate the problem of detecting lost tokens where the deposits are made by mistake, and the AMM refuses to return the tokens. We propose threat indicators by assuming the victim accounts' rational behavior after lost tokens. Our measurement result on Ethereum confirms the existence of lost tokens on Uniswap V3, Curve and Balancer (inflicting a total value of $57,000$ USD lost unfairly by these AMMs). We note that unfairness could imply other threats than theft and lost tokens, and a comprehensive study is left for future research. 

{
By design, all six AMM services measured are vulnerable to either token theft or lost tokens. The root cause is two-fold: First, the ERC20 standard has design flaws. Specifically, the \texttt{transfer} function does not allow tracking the history of token senders, rendering it hard or costly to enforce access control. Second, token implementations may exhibit the behavior undefined in the ERC20 standard, such as token rebase. This undefined behavior causes token theft (see \S~\ref{sec:counter:space} in detail). 
Besides, the discovered design flaws of ERC20 standard apply to any DeFi smart contracts built on ERC20 tokens, beyond just DEX/AMM (discussed in \S~\ref{sec:riskanalysis}). 
}

\vspace{2pt}\noindent\textbf{Mitigation}. 
We propose a secure design of AMM pool, in which the pool verifies the deposit transaction using ETHRelay protocols~\cite{me:ethrelay,DBLP:conf/blockchain2/FrauenthalerSSS20,me:btcrelay,DBLP:conf/aft/DaveasKKZ20} and can enforce access control on withdrawal against the original depositor account. While avoiding thefts and lost tokens, the ETHRelay procedure in smart contracts incurs a high cost, optimizing which is left for future works toward practicality. In addition to redesigning AMM pools, we propose building an off-chain infrastructure (a benign bot) to automatically discover withdrawal opportunities, claim the value ahead of actual attackers, and refund the victims. 

{
Our secure redesigns of AMM pool for Uniswap V2/V3 are open-sourced at GitHub~\cite{me:mitigate:v2}/~\cite{me:mitigate:v3}.
}

{
Note that existing ``fair'' or fair exchange protocols~\cite{DBLP:conf/ccs/CampanelliGGN17,DBLP:journals/corr/abs-1911-09148,DBLP:conf/ndss/MalavoltaMSKM19,DBLP:conf/ccs/0001MM19,DBLP:journals/corr/MillerBKM17,DBLP:conf/podc/Herlihy18,DBLP:conf/ccs/CampanelliGGN17} do not help solve the security risks uncovered in this work. The latter stems from the ERC20 token standard and is applicable to any DeFi smart contracts built on ERC20 tokens. As validated in our study (see \S~\ref{sec:rw}), implementing protocol-level fair exchanges on real-world ERC20 tokens faces the same implementation-level risk discovered in this work. 
}

\vspace{2pt}\noindent\textbf{Contributions}. Our contributions are outlined below: 

\vspace{2pt}\noindent$\bullet$\textit{ New findings}. We discover a total of $3.88$ million USD worth of stolen value by measuring and analyzing unfair trades on six popular AMM services on Ethereum and BSC. 
The discovered token theft attacks uncover the previously unknown patterns of extractable value (e.g., token rebase, interest, airdrop, buggy application smart contracts) and expose the adaptive attack strategies that occur offline and otherwise would have been hidden. 
{ After disclosure to both AMM and token developers, the discovered bugs have been confirmed by the Uniswap team.}

\vspace{2pt}\noindent$\bullet$\textit{ New techniques}. We develop an end-to-end analytical pipeline to detect unfairness and relevant attacks from Ethereum transaction traces, including an efficient join algorithm and various heuristics to automatically discover and cross-check token theft and lost tokens. 

\noindent
{\bf Roadmap}: We survey the related work in Section \S~\ref{sec:rw}. Section \S~\ref{sec:background} introduces the background of AMM and its specification. The discovery of unfair trades is in \S~\ref{sec:detectviolatingswap}. Detection of theft attacks is presented in \S~\ref{sec:attack}. Detection of lost tokens is in \S~\ref{sec:losttokens}. 
Countermeasures are presented in \S~\ref{sec:mitigate}. Bug disclosure is in \S~\ref{sec:disclosure} with the conclusion in \S~\ref{sec:conclude}.

\section{Related Work}
\label{sec:rw}

We survey the most relevant works and compare them against this one. Due to the space limit, we defer the extended related work to Appendix~\ref{appdx:rw}.

\noindent
{\bf MEV}: In DeFi, a prominent class of attacks exploit transaction reordering to gain blockchain (or maximal) extractable value or MEV (also denoted by BEV in the literature). At an abstract level, a MEV can be extracted by two essential steps: An adversarial account first looks for {\it profitable opportunity} and then, to claim the profit, sends {\it frontrunning/backrunning} transactions against competing accounts in a so-called Priority Gas Auction (PGA) game. There could be unlimited causes of profitability opportunities in smart contracts, and the notable ones are arbitrage, sandwich attacks, liquidation, etc. 

Existing research detects MEV attack instances from Ethereum transactions. As shown in Table~\ref{tab:bev}, the research works can be classified by the profitable opportunities exploited and the attack vectors (i.e., frontrunning, backrunning, or others). Specifically, Damian, et al.~\cite{DBLP:conf/sp/DaianGKLZBBJ20} are among the first to study extractable value on Ethereum and detect the arbitrage-by-frontrunning incidents on early-day DEX'es (e.g., TokenStores). The detection method is by collecting blockchain data for confirmed transactions and monitoring mempools for unconfirmed transactions. Torres, et al.~\cite{DBLP:conf/uss/TorresCS21} present a large-scale measurement study on Ethereum that detect generic frontrunning attacks by finding pairs of transactions (that win and lose the frontrunning game). Qin, et al.~\cite{DBLP:journals/corr/abs-2101-05511} measure MEV attacks on Ethereum of various causes including liquidation, sandwich attacks, arbitrage, and others captured by their transaction-replay techniques. Wang, et al.~\cite{DBLP:journals/corr/abs-2105-02784} detect the instances of arbitrage on Uniswap-V2, by finding profitable cycles on the token-transfer graph. Zhou, et al.~\cite{DBLP:conf/sp/ZhouQCLG21} discover arbitrage opportunities by finding profitable cycles among token-token exchange rates in real time. Xia, et al.~\cite{DBLP:journals/pomacs/XiaWGSYLZXX21} detect scam tokens listed on Uniswap V2 that impersonate other tokens to trick their traders; the scam-token signature is the pool of a shorter-lived life span of transactions than normal pools.

{\it This work's distinction to known MEV}: Arbitrage and sandwich attacks are two known patterns to extract MEV from AMM. In arbitrage, the profitable opportunities come from the difference in token exchange rates between AMM pools. In sandwich attacks, the profitable opportunities come from the imminent price changes caused by pending trades. 1) Both MEV attacks exploit or manipulate token prices. In both attacks, the attacker monitors certain conditions on the token prices to determine the timing of value extraction.
In this work, the ``extractable value'' stolen from the theft comes from the AMM design flaw that allows permissionless withdrawal. To extract the value, the attacker does not monitor token prices; instead, she monitors the presence of exploitable deposit (non-standard token deposit such as rebase) to determine when to extract the value.
2) In addition, in existing arbitrage and sandwich attacks, the attack success does not require breaking the fairness of individual AMM operations. That is, a successful sandwich attack (or arbitrage attack) can be composed of a sequence of token swaps that are fair. By contrast, the theft attacks studied in this work require breaking the fairness of AMM operations.
Therefore, the theft attacks studied in this work are fundamentally different from the MEV-based attacks~\cite{DBLP:conf/uss/TorresCS21,DBLP:journals/corr/abs-2101-05511,DBLP:journals/corr/abs-2105-02784,DBLP:conf/sp/ZhouQCLG21}, due to different causes and execution conditions. 

\begin{table}[!htbp] 
\caption{Related works on discovering and measuring MEV.}
\label{tab:bev}\centering{\small
\begin{tabularx}{0.5\textwidth}{ |l|X| }
  \hline
 Research works  & MEV causes \\ \hline
 Damian, et al.~\cite{DBLP:conf/sp/DaianGKLZBBJ20}
 & Arbitrage (multi-trade) \\ \hline
 Torres, et al.~\cite{DBLP:conf/uss/TorresCS21}
 & Generic (multi-trade) \\ \hline
 Qin, et al.~\cite{DBLP:journals/corr/abs-2101-05511} 
 & Arbitrage, sandwich, liquidation, replayable txs (multi-trade) \\ \hline
 Wang, et al.~\cite{DBLP:journals/corr/abs-2105-02784} 
 & Arbitrage (multi-trade) \\ \hline
 Zhou, et al.~\cite{DBLP:conf/sp/ZhouQCLG21} 
 & New arbitrage (multi-trade) \\ \hline
 This work
 & Permissionless pool withdrawal (single-trade) \\ \hline
\end{tabularx}
}
\end{table}

\ignore{
\begin{table}[!htbp] 
\caption{Related works on discovering and measuring MEV.}
\label{tab:bev}\centering{\small
\begin{tabularx}{0.5\textwidth}{ |l|X|X| }
  \hline
 Research works  & MEV causes & Attack vectors \\ \hline
 Damian, et al.~\cite{DBLP:conf/sp/DaianGKLZBBJ20}
 & Arbitrage (multi-trade) & Frontrun \\ \hline
 Torres, et al.~\cite{DBLP:conf/uss/TorresCS21}
 & Generic (multi-trade) & Frontrun (external) \\ \hline
 Qin, et al.~\cite{DBLP:journals/corr/abs-2101-05511} 
 & Arbitrage, sandwich, liquidation, replayable txs (multi-trade) & Frontrun/backrun \\ \hline
 Wang, et al.~\cite{DBLP:journals/corr/abs-2105-02784} 
 & Arbitrage (multi-trade) & N/A \\ \hline
 Zhou, et al.~\cite{DBLP:conf/sp/ZhouQCLG21} 
 & New arbitrage (multi-trade) & N/A \\ \hline
 This work
 & Permissionless pool withdrawal (single-trade) & Frontrun/backrun \\ \hline
\end{tabularx}
}
\end{table}
}

{
\noindent{\bf 
Provable secure swap protocols}: Protocol designs for fair swaps have been studied extensively in the literature. Different swap models have been considered including contingent payments~\cite{DBLP:conf/ccs/CampanelliGGN17} in which swaps occur between an off-chain digital good and an on-chain token, payment channels/networks~\cite{DBLP:journals/corr/abs-1911-09148,DBLP:conf/ndss/MalavoltaMSKM19,DBLP:conf/ccs/0001MM19,DBLP:journals/corr/MillerBKM17} in which multiple swaps occur off-chain and are batched into fewer transactions on-chain, fair cross-chain swaps~\cite{DBLP:conf/podc/Herlihy18} in which pegged tokens on one blockchain are swapped with tokens on another blockchain, etc. Most existing protocols consider the model of swapping assets between two mutually untrusted blockchain accounts (e.g., EOA in Ethereum) mediated through smart contracts or scripts on blockchain. Internally, they are constructed using primitives such as zero-knowledge proofs (e.g., in ZKCP~\cite{DBLP:conf/ccs/CampanelliGGN17}), time-lock smart contracts (e.g., in payment channels and fair cross-chain swaps). 

This work focuses on the case of deployed AMMes on Ethereum, and they should be differentiated from the provable secure swaps in the following senses: First, AMM swaps assets between an EOA and a smart-contract account (i.e., a pool), while the provable secure protocols consider swapping between two EOAs with counterparty risk. Second, AMM swaps two on-chain tokens, which are different from the swap models described above. 
}

\noindent
{\bf Smart contract vulnerabilities and attacks}: Existing research has covered the automatic discovery of new smart-contract vulnerabilities~\cite{DBLP:conf/ccs/LuuCOSH16,DBLP:conf/uss/KruppR18} and the detection of attack instances of known vulnerabilities~\cite{DBLP:conf/uss/ZhangZZL20,DBLP:conf/uss/ZhouYXCY020,DBLP:conf/uss/SuSDL0XL21,DBLP:conf/uss/0001L21}. These two lines of researches tackle generic program-level attacks such as reentrancy attacks~\cite{DBLP:conf/ndss/RodlerLKD19,DBLP:conf/sp/CecchettiYNM21} and transaction ordering dependencies, which are not DeFi specific as in this work.

{
\noindent{\bf 
Blockchain denial of service}: The system service of a public blockchain can be denied by manipulating mining incentives~\cite{mirkin2019bdos}, flooding the network with spam transactions~\cite{DBLP:conf/fc/BaqerHMW16,DBLP:conf/ccs/LiWT21,DBLP:conf/imc/LiTCWL21}, creating Sybil nodes to partition the network and eclipse victim nodes~\cite{DBLP:conf/uss/HeilmanKZG15}, and exhausting nodes' computing resource by running computation-intensive smart contracts~\cite{me:doers21,DBLP:conf/ndss/0002L20}. Recently proposed DETER attacks show that one can send invalid transactions at zero Ether cost to deny the mempool service~\cite{DBLP:conf/ccs/LiWT21,DBLP:conf/imc/LiTCWL21}. Gas based protection can be evaded by crafting smart contracts with under-priced instructions~\cite{DBLP:conf/ndss/0002L20}.

The DoS in this work refers specifically to lost tokens, in which a user having deposited tokens is unable to withdraw value. Lost tokens need not run large computations or send a flood of many transactions. Gas and transaction fees do not protect Ethereum from lost tokens.
}

\section{Background}
\label{sec:background}

Ethereum blockchain supports two types of addresses: An externally owned account (EOA), which is an Ether owner's public key, and a contract account (CA), which is the memory address the smart contract runs on miners. An Ethereum transaction includes a sender address, a receiver address, the Ether value transferred, and the optional input data that specifies smart-contract calls.

ERC20 is a popular standard on token interfaces on Ethereum. The ERC20 functions include \texttt{transfer} which allows a token owner to transfer her tokens to another account, \texttt{approve} which allows an owner to delegate the spending of her tokens to another delegatee account, \texttt{transferFrom} that allows a delegatee account to spend tokens on behalf of the original owner under the approved token limit, \texttt{balanceOf} that allows anyone to get an account's token balance, etc.

In an AMM or automated market maker, there are four parties: an AMM pool (a deployed smart contract), two tokens, say $T_0$ and $T_1$ (deployed smart contracts), traders or liquidity providers (each of which owns an externally-owned account/EOA or a deployed smart contract account). 
An AMM minimally supports two operations: 1) token swap between a trader account and the pool smart contract, and 2) liquidity addition/removal between a provider account and the pool smart contract. 
A trader (liquidity provider) can be either a normal user or a malicious attacker. 

In a $swapToken$ operation, the trader deposits $dx$ units of token $T_0$ to the pool's account and receives from the pool $dy$ units of token $T_1$. The AMM internally determines the token exchange rate (or value of $dy$) by enforcing the invariant on some function. For instance, Uniswap is a constant-product market maker where the invariant function is the product of the two token balance.

In an $addLiquidity$ operation, the liquidity provider deposits $dx$ units of token $T_0$, deposits $dy$ units of token $T_1$, and withdraws a certain amount of ``base'' tokens $T_{\texttt{LP}}$. A remove-liquidity workflow is the reverse of add-liquidity, in that the liquidity provider deposits tokens $T_{\texttt{LP}}$ and withdraws tokens $T_0$ and $T_1$ at a certain ratio.

AMM protocols instantiate the above specification. Token deposit can be realized by either directly \texttt{transfer} the tokens to the pool or doing so indirectly via calling \texttt{approve} and \texttt{transferFrom} functions. Different protocols' functions are listed in Table~\ref{tab:uniswapv2:api}.

\begin{table}[!htbp] 
\caption{AMM protocols' API in $swapToken$ and $addLiquidity$ operations. $T_0$/$p$/$*$ represent deposited token/pool/arbitrary smart contract.}
\centering{\footnotesize
\begin{tabularx}{0.5\textwidth}{ |l|X|l| } 
  \hline
Layer & $swapToken$ & $addLiquidity$ \\ \hline
Uniswap V2
& $T_0$.\texttt{transfer}() & $T_0$.\texttt{transfer}() \\ 
& $p$.\texttt{swap}() & $T_1$.\texttt{transfer}() \\ 
 & & $p$.\texttt{mint}() \\ 
\cline{2-3}
& $*$.\texttt{transferFrom}() & $*$.\texttt{transferFrom}() \\ 
& $p$.\texttt{swap}() & $*$.\texttt{transferFrom}() \\ 
 & & $p$.\texttt{mint}() \\ 
\hline
Uniswap V3
& $p$.\texttt{swap}() & $p$.\texttt{mint}() \\ 
\hline
Balancer
& $p$.\texttt{swapExactAmountIn}() & $p$.\texttt{joinpool}() \\ 
\hline
Curve
& $p$.\texttt{exchange}() & $p$.\texttt{add\_liquidity}() \\ 
\hline
\end{tabularx}
}
\label{tab:uniswapv2:api}
\end{table}

\noindent
{\bf Fairness specification}: 
An AMM operation executes a sequence of token deposits and withdrawals; the correctness means fair operation execution. Without loss of generality, we use token swap as an example in the description, while the extension to liquidity operations is straightforward. Given a token swap, we call the account calling the deposit function by the depositor and the account receiving the token withdrawal by the withdrawer.

\begin{figure}[!ht]
\begin{center}
\includegraphics[width=0.5\textwidth]{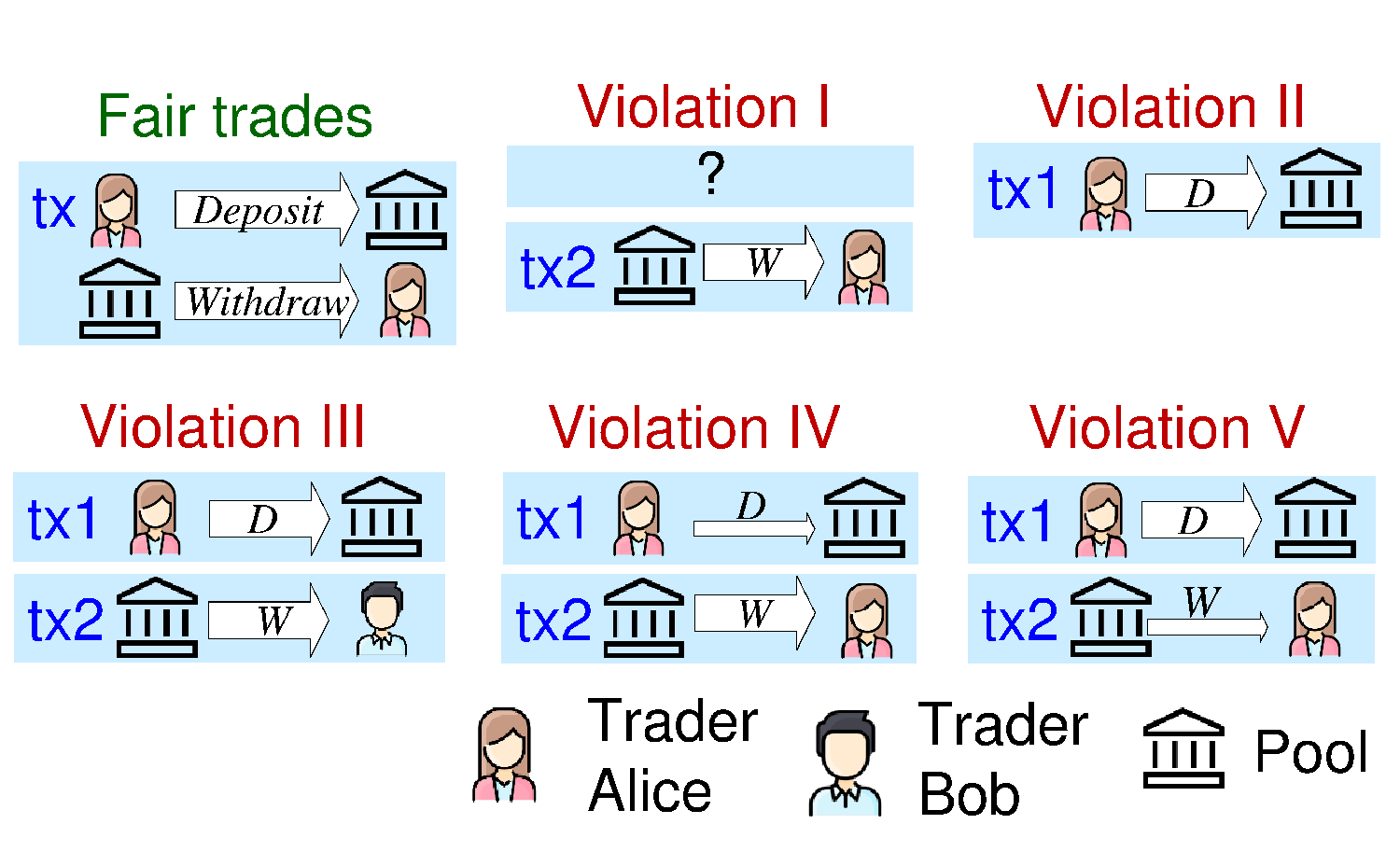}
\end{center}
\vspace{-0.1in}
\caption{Fair trades and violation cases.}
\label{fig:swaps}
\end{figure}

\begin{itemize}[leftmargin=0.5cm]
\item
A fair token swap consists of a token deposit of one token worth value $v$ and a withdrawal of another token worth the same value. The depositor and withdrawer accounts are either the same or match a function call if both deposit and withdrawal are in the same transaction. 

\item
{\bf Fairness violation I} is a standalone withdrawal that does not match any token deposit.

\item
{\bf Fairness violation II} is a standalone deposit that does not match any token withdrawal. 

\item
{\bf Fairness violation III} is an account mismatch in which the depositor and withdrawer accounts in a sequence of otherwise matched deposits and withdrawals do not match. For instance, as shown in Figure~\ref{fig:swaps}, the deposit is issued from Account Alice, and the withdrawal is from Bob.

\item
{\bf Fairness violation IV} is a lower-value deposit in which the deposited value is lower than the value withdrawn.

\item
{\bf Fairness violation V} is a higher-value deposit in which the deposited value is higher than the value withdrawn. Both Violations IV and V are cases of value mismatch.
\end{itemize}

Figure~\ref{fig:swaps} shows a fair trade and five patterns violating trade fairness.
{%
In particular, we stress that Violations IV/V are different from the existing extractable-value (MEV) attacks. In our violation cases, each instance deals with a {\it single} AMM operation, say a swap, and a single exchange-rate value. That is, it is a pair of token deposit and withdrawal where the exchange rate takes effect upon the withdrawal. Suppose in a swap, Alice deposits $x$ units of token $T_0$ and then withdraws $y'$ units of token $T_1$. If it is a violation IV, $y'>y$ where $x/y$ is the exchange rate upon withdrawal. This is unlike the existing MEV attacks, where each attack instance deals with {\it multiple} AMM operations and exploits the change of exchange rates at different operation times, such as arbitrage and sandwich attacks.
}


At last, we formulate two AMM risks: token theft attacks and lost tokens as a risk of losing usability.

\noindent
{\bf Threat: Theft attack}: An attacker trader monitors a target AMM pool's state, both on and off the blockchain. The attacker estimates the profitability of the AMM pool and, upon the right timing, sends crafted transactions to call AMM operations. The attack/theft is successful if the attacker's account is able to withdraw a value higher than the value she deposited. Specifically, the attacker's capabilities include monitoring confirmed transactions in the blocks and unconfirmed transactions in mempools. 

The victim can be the AMM pool itself or another account from which the attacker extracts deposited value. In the latter case, the victim account can be another trader (e.g., in Violation III) or another non-standard ``depositor'' (e.g., in Violation I). For instance, a non-standard depositor can be a token issuer rebasing the token supply (i.e., $P_1$ as described in \S~\ref{sec:profitabledeposits}).

In addition, we formulate the notion of ``lost tokens'', which indicates the lack of desirable features, or in other words, poor usability.

\noindent
{\bf Poor usability: Lost tokens}: Suppose a trader deposits a token to an AMM pool using an unsupported interface. The trader who wants a refund of the token can not succeed. This results in an unwanted situation where the token or Ether deposited by mistake is permanently lost.


\section{Uncovering Unfair Trades}
\label{sec:detectviolatingswap}
\label{sec:mmalgo}

\noindent
{\bf Input data}:
Our input data is raw blockchain transactions, including function calls and log events by smart contracts. We obtain the transaction data by crawling Ethereum exploration services including \url{Etherscan.io}~\cite{me:etherscan}, \url{BSCscan.io}~\cite{me:bscscan} and relevant BigQuery datasets~\cite{me:bigquery:ethereumdataset}. 
We only use the event log emitted from the AMM pool smart contract (which we assume is trusted), such as Uniswap V2's pool.
From the raw input data, we attribute the \texttt{transfer}/\texttt{transferFrom} calls to token deposits and withdrawals as described. Besides, we collect the invocations to relevant functions, such as \texttt{swap} in Uniswap V3 in Table~\ref{tab:uniswapv2:api}; from there, we can filter out the deposits and withdrawals that are mapped to fair operations. 
We also use the pool's events to reconstruct the mapping between deposit value and withdrawal value in the withdrawal record.

\noindent
{\bf Design challenge}:
A naive way of finding unfair trades is to filter out fair operations by extensively joining token deposits with withdrawals. Because a fair AMM operation can contain an arbitrary number of deposits and withdrawals, the naive join needs to be built on the {\it powerset} of token deposits and withdrawals, which is extremely expensive and unscalable (Note that the powerset of $n$ deposits/withdrawals is of $2^n$ elements).
To avoid inefficiency, our observation is that due to Ethereum's single-threaded execution model, a token deposit can be matched only to a subsequent withdrawal before the next deposit occurs. This property allows to join token deposits and withdrawals based on their timings and makes it possible for linear-time search.

\noindent
{\bf The algorithm} takes as input a list of function calls representing token deposits and withdrawals. In the first round, the algorithm equi-joins the function-call records by their transaction IDs. If the deposits and withdrawals in the same transaction also match in value, it emits them as a fair operation before removing them from the list. If the value does not match, the algorithm merges the intra-transaction deposits and withdrawals into one virtual record (e.g., a deposit of value equal to the originally deposited value deducted by the withdrawn value).

In the second round, the algorithm joins deposits and withdrawals across transactions. It linearly scans the list of deposits and withdrawals sorted by time: For two consecutive deposits, say $d1$ and $d2$, it finds any subset of all withdrawals between $d1$ and $d2$ that matches $d1$ in value. For each match, the algorithm removes the deposit and withdrawals from the list before emitting a fair operation. The algorithm then reverses the process; that is, for two consecutive withdrawals, it finds a set of deposits between the two withdrawals that match the earlier withdrawal in value.

In the third round, the algorithm linearly scans the remaining list and emits the token deposits and withdrawals as either value mismatch or standalone operations. The pseudocode is in Appendix~\ref{sec:algorithm:detect}.

\begin{figure}[!ht]
\centering
\includegraphics[width=0.29\textwidth]{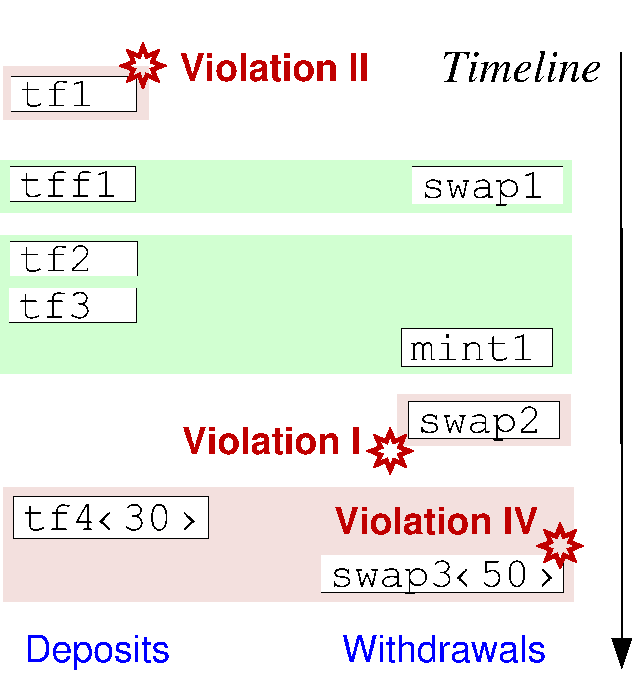}
\caption{Matchmaking deposits and withdrawals: Matched cases are in green boxes, and mismatches are rendered in red, including standalone withdrawal (I), standalone deposits (II), and deposits of lower value (IV). In the figure, \texttt{tf}/\texttt{tff} refer to functions \texttt{transfer}/\texttt{transferFrom}.}
\vspace{-0.1in}
\label{fig:joinproblem}
\end{figure}

\noindent
{\bf Example}: We show an example of running the unfairness discovery algorithm in Uniswap V2. Uniswap V2 supports token deposits by both \texttt{transfer} and \texttt{transferFrom}. It also supports token withdrawals by \texttt{swap} (in $swapToken$) and by \texttt{mint} (in $addLiquidity$). We collect the corresponding function-call records for token deposits and withdrawals. Figure~\ref{fig:joinproblem} shows the collected deposits and withdrawals, ordered by time. 
In the first round, the algorithm identifies and filters out fair operations by linking deposits and withdrawals through transaction ID, which is \texttt{tff1}-\texttt{swap1}. In the second round, it then matches token deposits and withdrawals by timing. For instance, \texttt{tf2} and \texttt{tf3} are matched to \texttt{mint1} by value, which leads to a fair $addLiquidity$ operation. In the last round, the algorithm emits the mismatches in the remaining list. Specifically, \texttt{tf4} is mapped to \texttt{swap3}, but their value does not match, thus producing a case of Violation IV. This leaves two standalone operations in the list, namely, \texttt{tf1} (Violation II) and \texttt{swap2} (I).


\begin{table}[!htbp] 
\caption{Discovered unfair trades on top DEXes: value is presented in the unit of million USD, and the number of transactions violating fairness (i.e., \#tx) is presented in the unit of a thousand.}
\label{tab:swapsviolating}\centering{\scriptsize
\begin{tabularx}{0.5\textwidth}{ lXlllllX } 
\hline
 & Fair ($10^6$) & I & II & III & IV & V & \#tx ($10^3$) \\ \hline
Uniswap-V2 & $618$ & {\color{red} \bf $1.26$} & $.36$ & {\bf$.49$} & $.09$ & {\color{red} \bf $3.64$} & $212$ \\ \hline
Sushiswap & $.17$ & {$\approx{}0$} & {$.03$} & {\bf$.06$} & $\approx{}0$ & $10^{-3}$ & $.33$ \\ \hline
Pancakeswap & $73$ & {$.53$} & $.12$ & {\bf$.05$} & $.99$ & $0.15$ & $452$ \\ \hline
Uniswap-V3 & $.053$ & $0$ & {$.17$} & $0$ & $0$ & $0.04$ & $.17$\\  \hline
Balancer
& $.013$ & $0$ & {$.74$} & $0$ & $0$ & $0$ & $6.9$ \\ \hline
Curve
& $.13$ & $0$ & $\approx{}0$ & $0$ & $0$ & $0$ & $\approx{}0$\\ \hline 
\end{tabularx}
}
\end{table}

\noindent
{\bf Results}: We conduct the measurement and transaction analysis on all six DEX services.
In particular, we set the minimal $Z\%$ value difference to determine match/mismatch. If a deposit (withdrawal) has $Z\%$ higher value than the paired withdrawal (deposit), it is considered a mismatched case (IV/V). A rule of thumb is to set $Z\%=10\%$ on our dataset, which helps remove the negative cases, such as fees charged by tokens and router smart contracts.
For each DEX, we report the number of swaps and their total value under each category. 
To estimate the value in a violating operation $tx$, we use the exchange rate between the token $T$ and one of the three USD-pegged stablecoins (i.e., $USDT$, $USDC$, and $DAI$). Specifically, we find from the three Uniswap-V2 pools (i.e., $T$-$USDC$, $T$-$USDT$, and $T$-$DAI$) the swap transaction say $tx'$ that occurs at the nearest time to $tx$. We use the exchange rate in $tx'$ to estimate the value of token $T$ in transaction $tx$.
 
The results are in Table~\ref{tab:swapsviolating}, in which the majority of swaps, about $97.9\%$, are fair. Among the rest that violates fairness, there are two significant sources (in red): Violation I and V in Uniswap V2. For instance, cases violating I in Uniswap V2 are worth $1.2$ million USD. Cases violating V in Uniswap V2 are worth $3.64$ million USD. 

To explain these violations, we manually inspect the transactions. Cases violating V are caused by the high fees charged by token smart contracts (note that the pool smart contracts are provided by the AMM and their fees are normally standard). Existing research~\cite{DBLP:conf/ccs/0002ZLLWCXZ19} characterizes ERC20 tokens' fees and adversarial behavior in hiding the high fees from customers.

Cases violating I, that is, standalone withdrawals, are caused by non-standard token deposits. Specifically, when we collect transactions, we only consider the ``standard'' token deposits explicitly supported by the Uniswap V2 protocol. However, in practice, various non-standard token operations that update balance (e.g., supply rebase and token interest, as will be discussed in \S~\ref{sec:profitabledeposits}) are mistreated by the Uniswap V2 as a token deposit. Token withdrawals that occur after non-standard deposits can execute successfully. These withdrawals manifest as standalone withdrawals (i.e., Violation I).

{%

\subsection{Patterns of Violations}
\label{sec:profitabledeposits}

We describe some profitable deposits found in our measurement result. We use Uniswap V2 as an example to show why such deposits are profitable and can be stolen in theft.

\begin{figure}[!htbp]
\begin{center}
\includegraphics[width=0.475\textwidth]{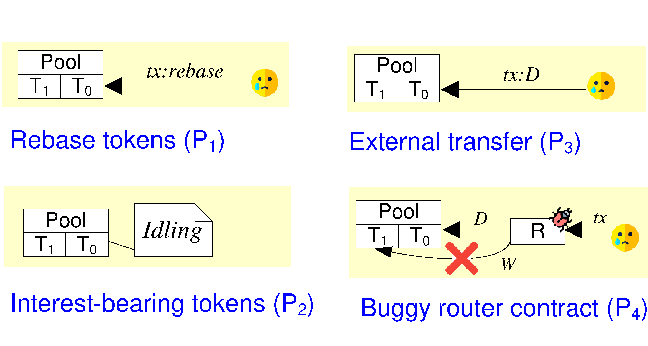}
\end{center}
\caption{Patterns of profitable deposits: $R$ is the router smart contract (part of the pool), $T_0$/$T_1$ are the two tokens in the target liquidity pool. Both $P_1$ and $P_2$ are non-standard mechanisms that update token balance. $W$/$D$ refer to withdrawal/deposit (directly via calling \texttt{transfer}).}
\vspace{-0.2in}
\label{fig:patterns}
\end{figure}

\subsubsection{Token Supply Rebase (Pattern $P_1$)}
\label{sec:cases:rebase}

In many real-world tokens, the account balance can be changed by non-standard operations beyond token transfer. We describe an example of {\it token rebase}. Token balance can be elastically adjusted by token issuers for purposes such as stabilizing the token's exchange rates against other tokens. In practice, token rebase is such a non-standard balance-changing function, supported in some widely used tokens, such as AMPL~\cite{me:ampl:token}. When a token, say $T_0$, is rebased, a Uniswap's relevant pool, say for tokens $T_0$ and $T_1$, could have its $T_0$ balance increased. Uniswap falsely assumes that this balance increase is due to some accounts that swap token $T_0$ for token $T_1$. Uniswap's pool would mistakenly admit anyone who requests to withdraw token $T_1$ after the rebase of $T_0$.

\ignore{
{\it Token shareholder}: 
As another example, there are tokens that charge ``tax'' to average owner accounts and redistribute the ``tax'' income among privileged accounts of shareholders. Specifically, these tokens support two kinds of owners, average owner, and token shareholders. When there is a transfer, the token smart contract charges a certain amount of fees. The fees are redistributed to the privileged accounts of token shareholders. Thus, a transfer between average accounts would update the balance of all shareholders' accounts. In a Uniswap's pool of token types $T_0$ and $T_1$, if the pool's account in $T_0$ is a shareholder, any transfers in $T_0$ would increase the pool balance and create a profitable opportunity.
 
Furthermore, even when the pool receives a token deposit, the deposit may not be intended as a part of the swap. For instance, airdrop is a common practice to promote the use of a new token type, in which token issuers periodically give away newly minted tokens to the current token owners (e.g., in Tokens MILK2~\cite{me:token:milk2} and 3Cs~\cite{me:token:3cs}). When a token, say $T_0$, is airdropped into a Uniswap's pool, the pool holding tokens $T_0$ and $T_1$ would mistakenly treat the transfer as a part of swap from token $T_0$ to $T_1$. Thus, the pool would admit anyone's request to withdraw token $T_1$ after airdrop of token $T_0$ to the pool's account.
}

In general, these non-standard token operations allow for updating pool balance but do not withdraw the same value from the pool. We name the pattern by $P_1$.

\subsubsection{Interest-bearing Tokens ($P_2$)}
Beyond $P_1$, there are implicit approaches to updating the balance. Many tokens are bearing interest. That is, the balance in the token smart contract increases at a certain rate as time goes by. Suppose a Uniswap pool is of token $T_0$ and $T_1$ where $T_0$ is interest bearing. Similar to $P_1$, the Uniswap pool observing the increase of token $T_0$ would admit the withdrawal of token $T_1$ by anyone who requests it, including a value-extracting attacker. We call this attacker's pattern by $P_2$, which is different from $P_1$. In $P_2$, the attacker simply waits long enough to harvest the extractable value from token interest. 

\subsubsection{External Transfer for Deposit ($P_3$)}

In Uniswap, a trader can externally call token $ T_0$'s transfer function to make a deposit so that she can withdraw token $T_1$ in a separate transaction. We call the pattern of a victim trader sending a transaction to externally transfer token $T_0$ by $P_3$. When $P_3$ occurs, a malicious account can send the withdrawal transaction ahead of the victim's one to claim the value in the pool. This attack is possible because a Uniswap pool receiving a deposit of $T_0$ will approve the next withdrawal of token $T_1$, no matter who sends it. 

\subsubsection{Buggy Router Contracts ($P_4$)}

In the above three patterns, the victim may externally call token $ T_0$'s functions. In the fourth pattern $P_4$, the trader delegates the handling of the swap to a router smart contract that is buggy and violates fairness. After the trader EOA's transaction, the router smart contract successfully completes the deposit of Token $T_0$ but fails at withdrawing Token $T_1$. Thus, anyone who requests the token withdrawal in the next transaction, including an attacker EOA, would succeed in extracting illicit value. 
}

\section{Detecting Token Thefts}
\label{sec:attack}

\subsection{The Detection Problem}
\label{sec:discoveryproblem}

\noindent{\bf Theft detection}: 
Our risk analysis (will be presented in \S~\ref{sec:riskanalysis}) shows the pool design flaws (i.e., vulnerability) that enable extractable value, that is, the value {\it can} be extracted by anyone on the Internet. However, whether the value is {\it actually} extracted by real-world attackers, or equivalently, whether the violation operations are actual attack instances, is a different problem. 
The detected fairness violations, especially I, III, and IV, could present the candidate cases for extracted value, but not (yet) the confirmed cases.

Consider as an example that a token issuer sends a vulnerable rebase transaction to a pool's token, followed by a withdrawal transaction sent by account $A$ (i.e., $P_1$). $A$ may or may not be the token issuer. The former case is not an attack, as the token issuer, aware of the risky rebase, sends the withdrawal to remedy the risk. The latter case may be an attack. In other situations, including $P_3$ and $P_4$, an unfair trade can also be attributed differently.
\ignore{
For instance, in $P_1$, after a token issuer sends an exploitable rebase transaction, it is equally likely one of the following cases occur: 1) the token issuer, aware of rebase being exploitable, quickly sends a swap request herself to claim the value in the Uniswap pool. 2) An attacker observing the occurrence of rebase operation calls the liquidity pool to withdraw the value there. The former case is not an attack, while the latter one is. In other cases, the same applies: In $P_3$, if the trader sends another transaction to withdraw the value in token $T_1$ right after the transaction of depositing $T_0$, it is a benign case. But if it is an attacker account that withdraws after the external deposit, it constitutes an attack. In $P_4$, if the trader observes the router exception and immediately send a ``make-up'' transaction to claim the value herself, it is a benign case. But if it is a third-party ill-intended account that sends the withdrawal transaction, it constitutes an attack.
}

Distinguishing attacks from benign cases entails checking two essential conditions: {\bf 1) Whether two accounts are controlled by two distinct physical users}: It cannot be an attack if the trader account who makes the profitable deposit in token $T_0$ is controlled by the same physical user with the account who succeed in withdrawing the value in token $T_1$. 
Thus, attack detection becomes the problem of linking two blockchain accounts of the same physical user, which is known to be an intractable problem if the only information known is on-chain data.
{\bf 2) What's a transaction sender's intention}: That is, whether a withdrawal transaction is intended as an attack. Even when the withdrawer and depositor accounts are not the same physical user, they can be friends in  real life, which makes the swap non-attack. Thus, it entails detecting the depositor account's intention in sending the withdrawal request, which, however, can be hidden opaque and pose challenges to detection.

\noindent
{\bf Feasibility of existing approaches}: 
On checking Condition 1), there are existing works that tackle the recovery of account linkage by heuristics specific to Bitcoin's UTXO model~\cite{DBLP:conf/socialcom/ReidH11,DBLP:conf/imc/MeiklejohnPJLMVS13}. The only work for Ethereum~\cite{DBLP:conf/fc/Victor20} presents a linkage heuristic exploiting some DEXes deposit addresses. Specifically, on order book-based DEXes (i.e., IDEX and EtherDelta), trader Alice swapping token $T_0$ for $T_1$ can specify the so-called deposit address who receives $T_1$. Assuming a swap is normal, the sharing of deposit addresses by two traders implies that the two traders are controlled by the same physical user. This work is different from our attack detection problem, where the deposit address heuristic is to detect the case that two accounts are linked, while our problem is to detect if two accounts are unlinked. More importantly, the deposit address heuristic works only under the assumption of normal swaps, while our work deals with fairness violations. Given a fairness-violating swap, the deposit address of a swap refers to the attacker. Thus, when an attacker targets two victim traders, it could appear that two traders share the same deposit address. In this attack, the two victim traders do not have to be linked to the same off-chain user.

On checking Condition 2), there are existing works on threat intelligence on blockchains by characterizing attackers' behavior~\cite{DBLP:conf/uss/SuSDL0XL21}. However, their targets are program-level attacks such as reentrancy attacks, and their attack strategies don't apply to the DEX context of this work.

\begin{figure}[!ht]
\begin{center}
\includegraphics[width=0.4\textwidth]{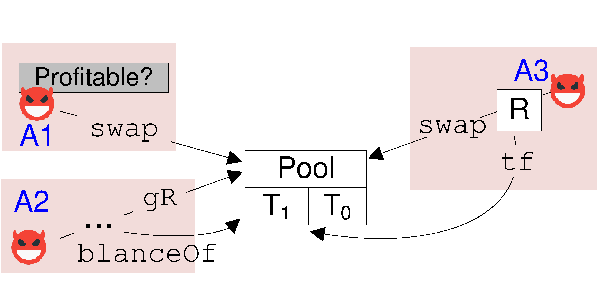}
\end{center}
\caption{Rational attack strategies (A1/A2/A3).}
\label{fig:indicators}
\end{figure}

\subsection{The Detection Method}

\noindent{\bf Detection design rationale}: We approach the theft detection problem and tackle the challenges above from three angles: 1) We propose various heuristics to model what a rational attacker account would do (i.e., $I_1,I_2,\dots,I_5$ as will be described). We then match the observable account behavior on-chain to the heuristics. For instance, a heuristic models aggressive accounts who eagerly send withdrawal upon observing an extractable deposit. 

However, given an extractable deposited value, an eager deposit could indicate, besides a malicious thief, a benign user (e.g., a pool safeguard) using an automated process to protect the depositor user's value. To distinguish the two possible cases, 2) we propose to study if there are frontrunning behavior, that is, multiple users sending transactions to withdraw the same value created by an extractable deposit. The idea is that {\it the frontrunning or the competition among multiple users indicates a conflict of interest among the users and that at least one user is ill-intended.} Otherwise, a single safeguard does not have the incentive to send multiple withdrawal transactions to frontrun itself. 

Furthermore, given the identified accounts, 3) we collect their tagged labels by searching forums and social networks on the Internet. The idea is that if an account is reported malicious, the withdrawal transaction from the account is more likely an attack.

Our overall detection method is to cross-check the three independent detection metrics on a candidate instance. We report positive instances if two detection metrics agree with each other.

\noindent{\bf
Method 1: Heuristics-based indicators}: Observing an extractable value created by a deposit, a rational thief would eagerly send the withdrawal transaction that is included in a block as close to the deposit transaction as possible. Based on the heuristic, we propose three indicators $I_1,I_2,I_3$ listed below. We defer the description of proposed theft-detection algorithm and additional indicators to Appendix~\ref{appd:extraindicators}.

\begin{center}\fbox{\parbox{0.90\linewidth}{
Indicator $I_1$: An account whose withdrawal transactions occur consistently within $X_1$ blocks after profitable deposits (i.e., block gap smaller than $X_1$) and whose victim deposits are not caused by token interests (i.e., not $P_2$) is likely to be an attacker. Formally, 
$$
I_1\leq{}X_1
$$
}}\end{center}

\begin{center}\fbox{\parbox{0.90\linewidth}{
Indicator $I_2$:
An account whose withdrawal transactions occur consistently more than $X_2$ blocks after profitable deposits (i.e., block gap larger than $X_2$) and whose victim deposits are based on token interests is likely to be an attacker. Formally, 
$$
I_2\geq{}X_2
$$
}}\end{center}

\begin{center}\fbox{\parbox{0.90\linewidth}{
Indicator $I_3$:
An account whose ratio between indicator $I_2$ and $I_1$ is larger than $X_3$ is likely an attacker. 
Intuitively, it shows the attacker's intentional adaptive strategies in exploiting deposits of different patterns.
Formally,
$$
\frac{I_2}{I_1}\geq{}X_3
$$
}}\end{center}

{%
An account who always checks the difference between token reserve in the pool smart contract and token balance in the token smart contract before token withdrawal from a pool is likely a general-purpose attacker.  
The idea is that if there is a difference between the return value from a pool's \texttt{getReserves()} function and that from token's \texttt{balanceOf()} function, it indicates the pool has extractable value. And an account's behavior to check the difference shows its intention to extract the value. 
We thus propose the following attack indicator.

\begin{center}\fbox{\parbox{0.90\linewidth}{
Indicator $I_4$: The attack signature is a sequence of two function calls 

Suppose $I_4\%$ of all the token withdrawals from an account are preceded by checking the difference between token reserve and balance, that is, calling functions \texttt{pool.getReserves()} and \texttt{erc20token.balanceOf()}. The account is an attacker, if the following condition is met.

$$I_4\geq{}X_4$$
}}\end{center}

Note that this indicator does not require the smart contract to actually withdraw the token $T_1$. 
}

\begin{center}\fbox{\parbox{0.90\linewidth}{
Indicator $I_5$: An account whose withdrawal transactions exclude any token deposit is likely to be an attacker.
}}\end{center}

\begin{center}\fbox{\parbox{0.90\linewidth}{
Indicator $I_6$:
Standalone withdrawal that extract value from deposit patterns $P_1$ and $P_2$ indicate an attacker, because the value in $P_1$ and $P_2$ is intended for AMM pools, not for external accounts. This intuition is materialized into the formula below:
$$
\frac{|P_1|+|P_2|}{\sum_{i}{|P_i|}}\geq{}X_6
$$
}}\end{center}

\noindent{\bf Method 2. Finding frontrunning}: Given each withdrawal involved in a violation, say $w$, we also detect if there are frontrunning withdrawals. We do so by scanning all blocks after the victim deposit and before the $10$th block after the withdrawal. Among these transactions and their internal calls, we find the ones that call the same pool account with $w$. We consider both successful and failed withdrawals. We report if a withdrawal violating fairness frontruns or is frontrun by at least one other withdrawal as indicator $I_7$.

\begin{center}\fbox{\parbox{0.90\linewidth}{
  Indicator $I_7$:
  An account who has one or more competitors is likely to be an attacker.
  }}\end{center}

\noindent{\bf
Method 3: External incident reports}:
We search the Internet sites, including Google, Twitter, and Reddit. Our searches include the candidate attacker accounts in Table~\ref{tab:attackerview} and keywords such as ``attack, vulnerable, scam, frontrunning, bot, arbitrager, MEV, bots.'' Besides, we check the candidate account's tags on Etherscan and if they are security sensitive.
Out of the $142$ suspected accounts from our results (described next in \S~\ref{sec:theftinstances}), we found $9$ accounts are complained about on the Internet, and $2$ accounts are tagged as MEV bots on Etherscan. 
The ground-truth accounts are presented in Table~\ref{tab:attackerview}, and the full detail of ground truth data collection is in Appendix~\ref{sec:groundtruth}.

{%
\begin{eqnarray}
\nonumber
&& I_1 \leq 2 (= X_1) \\ 
\nonumber
& \lor & I_2 \geq 617 (=X_2) \\
\nonumber
& \lor & {I_2}/{I_1} \geq 19.28 (=X_3)\\
\nonumber
& \lor & I_4 \geq 97.5\% (= X_4)\\
& \lor & \frac{\|P_1\|+\|P_2\|}{\sum_i{\|P_i\|}}\geq{}69.2\% (=X_6)
\label{eqn:parameters}
\end{eqnarray}
}

\ignore{
\begin{table}[!htbp] 
\caption{Detection parameters extracted from the ground-truth data}
\label{tab:parameters}\centering{
\begin{tabularx}{0.4\textwidth}{ Xlllll }
  \hline
  Parameters & $X_1$ & $X_2$ & $X_3$ & $X_4$ & $X_6$ \\ \hline
  Value & $2$ & $617$ & $19.28$ & $97.5\%$ & $69.2\%$ \\ \hline
\end{tabularx}
}
\end{table}
}

{
\noindent{\it Cross-validation}: We cross-validate the results detected under different methods. Specifically, in Method 3, we found $15$ accounts in online user complaints. We use the $15$ accounts as the ground truth to train the parameters in Method 1, namely $X_1,X_2\dots,X_6$. We obtain the trained Method 1 as in Equation~\ref{eqn:parameters}. 

\begin{table}[!htbp] 
\caption{Cross-validate results via different methods}
\label{tab:extractable_value_2}
\centering{
\begin{tabularx}{0.39\textwidth}{ |l|l|l|X| } 
\hline
Method 1 & Method 2 & Method 3 & { \# attackers} \\ \hline
* & * & * & 142 \\ \hline
\cmark & * & * & 139 \\ \hline
* & \cmark & * & 22 \\ \hline
* & * & \cmark & 15 \\ \hline
\cmark & \cmark & * & 22 \\ \hline
\end{tabularx}
}
\end{table}

We then apply the three methods separately on the $142$ candidate accounts detected earlier. Out of $142$ candidate accounts, $139$ accounts are detected positive under Method 1 with trained parameters. $22$ accounts are detected positive under Method 2. And $15$ accounts are found using Method 3, as shown in Table~\ref{tab:extractable_value_2}. {In summary, suppose we use the dataset of Method 3 as a ground truth and the dataset detected by matching Method 1 {\it or} Method 2. The false negative rate is $0\%$, and false positive rate unknown.}


}

\begin{table*}[h]
  \caption{Detected theft instances by accounts: Ground truth and test data. Bold are strong cases suggesting the account is an attacker. Victims include normal accounts and token issuers.}
\label{tab:attackerview}
  \centering{\scriptsize
  \begin{tabular}{p{6.5cm}p{0.6cm}p{1.9cm}p{0.4cm}p{0.4cm}p{0.4cm}p{1.4cm}p{0.2cm}p{0.5cm}p{0.3cm}}
  \hline
  Accounts & Value & {Deposit} & \multicolumn{7}{c}{Attack indicators} \\
  (Potential attacker) & $10^3$ \$ & patterns &$I_1$ & $I_2$ &$I_3$ & $I_4$ & $I_5$ &$I_6$&$I_7$\\ \hline
  \rowcolor{cyan!10}
  \multicolumn{10}{l}{\it Ground truth data} \\
  \hline
  $0x9799b475dec92bd99bbdd943013325c36157f383$ & 556 & 33$P_1$,4$P_2$,6$P_4$    &422       &371        &0.88  & 0                  &\xmark      &86\% &{\bf \cmark}\\
  \hline
  $0x0c08545df4939ef46c1364b5930e840739667467$ & 101 & 3$P_1$,1$P_4$            &{\bf 1.5} &           &0     & 0                  &\xmark      &75\% &\xmark\\
  \hline
  $0x56178a0d5f301baf6cf3e1cd53d9863437345bf9$ & 101 & 53$P_1$,4$P_2$,2$P_3$,2$P_4$ &13    & 1574      &{\bf121}   & 0                  &\xmark      &93\% &{\bf \cmark}\\
  \hline
  $0x0000000000007f150bd6f54c40a34d7c3d5e9f56$ & 11 & 15$P_2$                   &          &96         &$\infty$& 15 ({\bf 100\%}) &\xmark      &100\% &\xmark\\
  \hline
  $0x7c651d7084b4ba899391d2d4d5d3d47fff823351$ & 1.8 & 2$P_1$                   &68        &           &0     & 0                  &\xmark      &100\% &\xmark\\
  \hline
  $0x00000000e84f2bbdfb129ed6e495c7f879f3e634$ & 0.5 & 9$P_1$,4$P_3$            &{\bf 1.1} &           &0     & 0                  &{\bf \cmark}&69\% &\xmark\\
  \hline
  $0xca850b6833ef86ef0484c6be74b06d61b39df031$ & 0.5 & 40$P_3$                  &{\bf 1.4} &           &0     & 39 ({\bf 98\%})    &\xmark      &0\% &{\bf \cmark}\\
  \hline
  $0x17e8ca1b4798b97602895f63206afcd1fc90ca5f$ & 0 & 1$P_1$                     &3061      &           &0     & 0                  &{\bf \cmark}&100\% &\xmark\\
  \hline
  $0xf90e98f3d8dce44632e5020abf2e122e0f99dfab$ & 0 & 1$P_1$                     &1402      &           &0     & 0                  &\xmark      &100\% &\xmark\\
  \hline
  $0x42d0ba0223700dea8bca7983cc4bf0e000dee772$ & 0.5 & 1$P_3$                   &{\bf 0}   &           &0     & 1  ({\bf 100\%})   &{\bf \cmark}&0\% &\xmark\\
  \hline
  $0xEA674fdDe714fd979de3EdF0F56AA9716B898ec8$ & 0.2 & 2$P_1$,5$P_2$            &{\bf 0.5} &686        &{\bf1372}& 6 ({\bf 86\%})   &{\bf \cmark}&100\% &\xmark\\
  \hline
  $0x000000000025d4386f7fb58984cbe110aee3a4c4$ & 0.2 & 1$P_1$                   &38        &           &0     & 1  ({\bf 100\%})   &\xmark      &100\% &\xmark\\
  \hline
  $0xD224cA0c819e8E97ba0136B3b95ceFf503B79f53$ & 0.1 & 3$P_2$,1$P_3$            &32        &617  &19    & 4  ({\bf 100\%})   &\xmark      &75\% &\xmark\\
  \hline
  $0xb8aaed2e3117fa589eb05b0d7d0c8469d9e41ec5$ & 0.1 & 1097$P_1$                &351       &           &0     & 0                  &\xmark      &100\% &\xmark\\
  \hline
  $0x829BD824B016326A401d083B33D092293333A830$ & 0.1 & 5$P_2$                   &          &696  &$\infty$ & 5  ({\bf 100\%})   &\xmark      &100\% &\xmark\\
  \hline
  \rowcolor{cyan!10}
  \multicolumn{10}{l}{\it Test data} \\
  \hline
  $0x2a2e25ad00faac024f6a8d8cdc0fc698fbda71b7$ & 239 & $32P_1$                  &{\bf 0.03}&           &0     & 32 ({\bf 100\%})   &{\bf \cmark}&100\% &\xmark\\
  \hline
  $0xdb40ea5b6d0ef9e45c00c194345b44765a53dd19$ & 113 & $12P_1,2P_2,3P_3$        &{\bf 2.5} &665  &{\bf266}   & 17 ({\bf 100\%})   &{\bf \cmark}&82\%&{\bf \cmark}\\
  \hline
  $0xb6bf45f59b94d31af2b51a5547ef17ff81672743$ & 20 & $1091P_1,131P_2,9P_4$     &224       &600  &2.7   & 1231 ({\bf 100\%}) &{\bf \cmark}&99\%&{\bf \cmark} \\
  \hline
  $0xc762e15cfbf392f4346ee7e8466ca595e571382b$ & 16 & $64P_1,6P_2,13P_4$        &{\bf 3.0} & 5.3       &1.8   & 83 ({\bf 100\%})   &{\bf \cmark}&84\%&{\bf \cmark} \\
  \hline
  $0xa32d3f8a970ab3222ffa1c89cb910abf7b7201d0$ & 6 & $208P_2$                   &          &942 &$\infty$ & 208 ({\bf 100\%})  &{\bf \cmark}&100\%&\xmark \\
  \hline
\ignore{
}
\end{tabular}%
}
\end{table*}
  
\subsection{Results}

\subsubsection{Case Studies}
\label{sec:theftinstances}

We present one case study on two frontrunning withdrawals and two other cases on attacker accounts aggregating multiple withdrawals. The top accounts involved in unfair trades are characterized in Table~\ref{tab:attackerview}. The metrics include value stolen, causes, attacker indicators, the number of pools attacked, the number of victim accounts, and whether the account is an attacker. 

\begin{figure*}[!bhtp]
 \centering
 \includegraphics[width=0.8\textwidth]{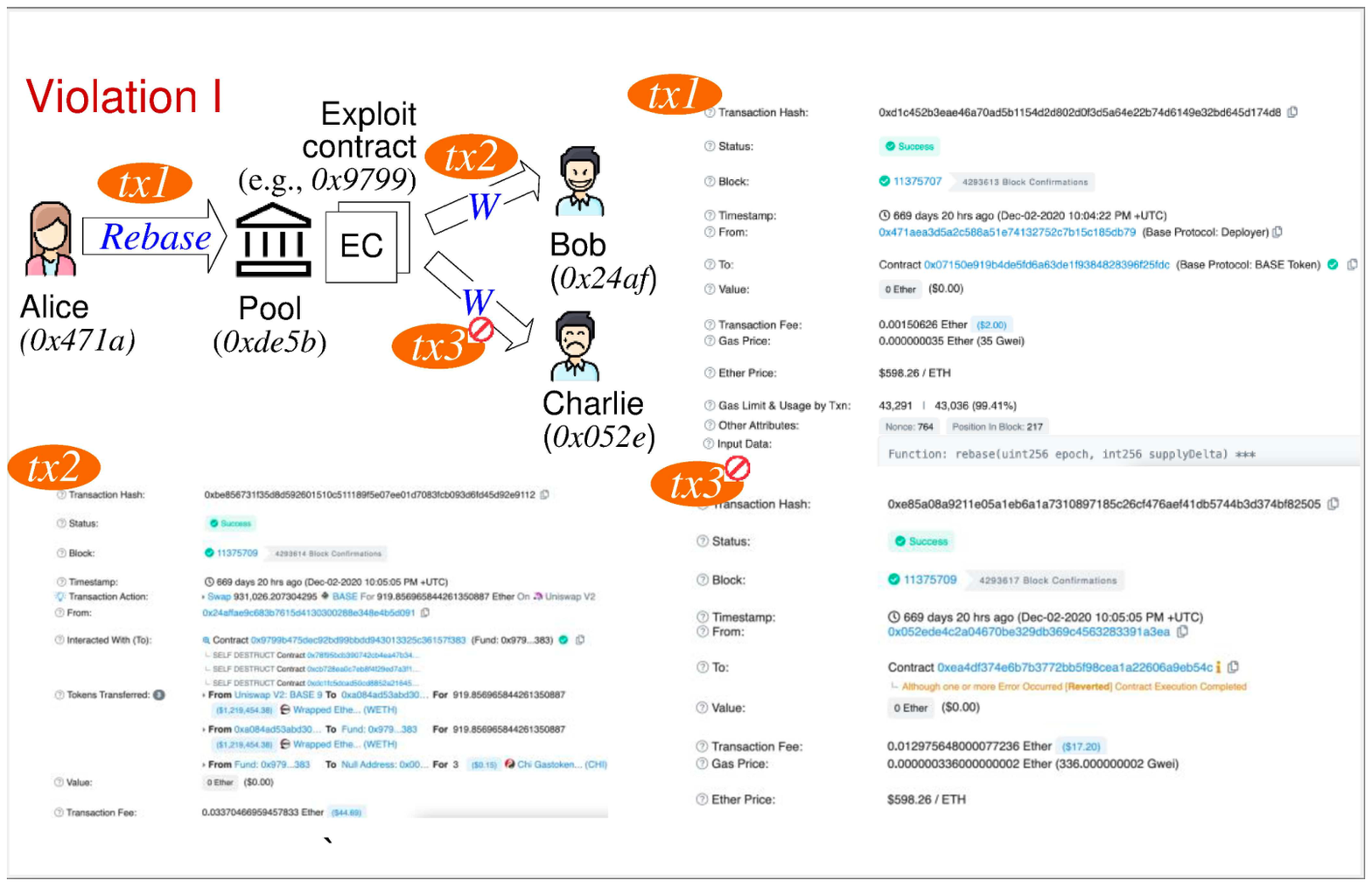} 
\vspace{-0.1in}
 \caption{Case study: A found trade violating fairness (I) shows frontrunning transactions ($tx_2$ and $tx_3$) to withdraw value from a Uniswap V2 pool created by token-rebase transaction $tx_1$. Transaction $tx_2$ with red cross mark is reverted.}
 \label{fig:casestudy}
\end{figure*}

\noindent{\bf Case 0. Frontrunning for claiming rebase}: This case is Violation I involving three transactions depicted in Figure~\ref{fig:casestudy}. First, BASE token issuer Alice (EOA $0x471a$) sends transaction $tx_1$ to rebase the token supply among all BASE token holders, including a Uniswap V2 pool account (CA $0xde5b$). $tx_1$ is included in Ethereum Block $\#11375707$. Then, two accounts, say Bob (EOA $0x24af$) and Charlie (EOA $0x052e$), respectively send transactions $tx_2$ and $tx_3$, to withdraw the extractable value created by the rebase. Both withdrawal transactions externally call some third-party smart contracts before calling the victim pool's ($0xde5b$'s) \texttt{swap()} function. For instance, Bob's withdrawal $tx_2$ externally calls the smart contract of account $0x9799$, which internally calls another smart contract of account $0xa084$, which further internally calls the pool's \texttt{swap} function. Likewise, Charlie's withdrawal $tx_3$ externally calls a smart contract $0xea4d$, which internally calls smart contract $01da$ before internally calling the pool's \texttt{swap} function. While both transactions are included in the same Block $\#11375709$, $tx_2$ is ordered before $tx_3$, and Bob successfully claims the extractable value, leaving Charlie's withdrawal transaction reverted. This whole process is depicted in Figure~\ref{fig:casestudy}.


\noindent
{\bf Case 1: Aggressive attacker}: Account $0x2a2e$ is involved in $32$ unfair trades and gains more than $2.39*10^5$ USD worth of illicit tokens. All $32$ swaps from this account are sent via a smart contract which checks the pool's profitability before withdrawal ($I_4$). In terms of timing, the malicious withdrawals are always sent in the same block with the risky, profitable deposit. We believe account $0x2a2e$ aggressively monitors the mempool for the target tokens' rebase transactions and if found, sends withdrawals immediately to ``backrun'' the deposits and claim the value.

\ignore{
\noindent
{\bf Case 2: General attacker (A2) w. off-chain monitoring}: Account $0x0c08$ has withdrawn four times and grabbed tokens of $1.01*10^5$ USD. The account directly and externally calls Uniswap pool's swap function to withdraw without checking pool profitability on-chain. The withdrawals are included in exactly one block after the one that includes the risky deposit. We believe Account $ 0x0c08$'s strategy is to monitor the pool profitability from an off-chain vantage point and to immediately send withdrawal by external calls. Because the account monitors the pool profitability instead of the cause (risky deposit), it has to wait until the inclusion of risky deposit that updates the pool state to be profitable. This explains why the inclusion of withdrawal has to be one block after the block including the risky deposit.
}

\noindent
{\bf Case 2: Attacker stealing interest}: Both Accounts $0xdb40$ and $0x5617$ have extracted value from interest-bearing tokens. Their block gaps are large ($665$ and $1574$, respectively). Given that there are enough samples, this consistent behavior shows both accounts' attack strategy to wait long enough for the token interest to accumulate and for the extractable value to be large enough before acting on it.

In general, for all accounts with indicators $I_1$ and $I_2$ in Table~\ref{tab:attackerview}, it is clear that the block gaps for different causes ($I_1$ and $I_2$) differ: The block gaps for non-interest causes are consistently much shorter than those for token interests. It implies that the attackers are aware of the situation (which causes them to exploit) and tailor their strategies to it.

\ignore{
\noindent
{\bf Case 5: Scavenger trader (A3)}: Account $0x9799$ is unlikely an intentional attacker: First, it does not read the token balance or check the pool's profitability before the withdrawal. Second, the block gap between the profitable deposit and malicious withdrawal is large, at hundreds of blocks apart. In addition, the smart contract Account $0x9799$ relays her withdrawal requests also make deposit to the pool. All these indicators show that $0x9799$ is a scavenger trader.
}

\subsubsection{Block Gaps}

\begin{figure}[!ht]
  \begin{center}
    \subfloat[Block gaps and the number of competing attackers]{%
\includegraphics[width=0.235\textwidth]{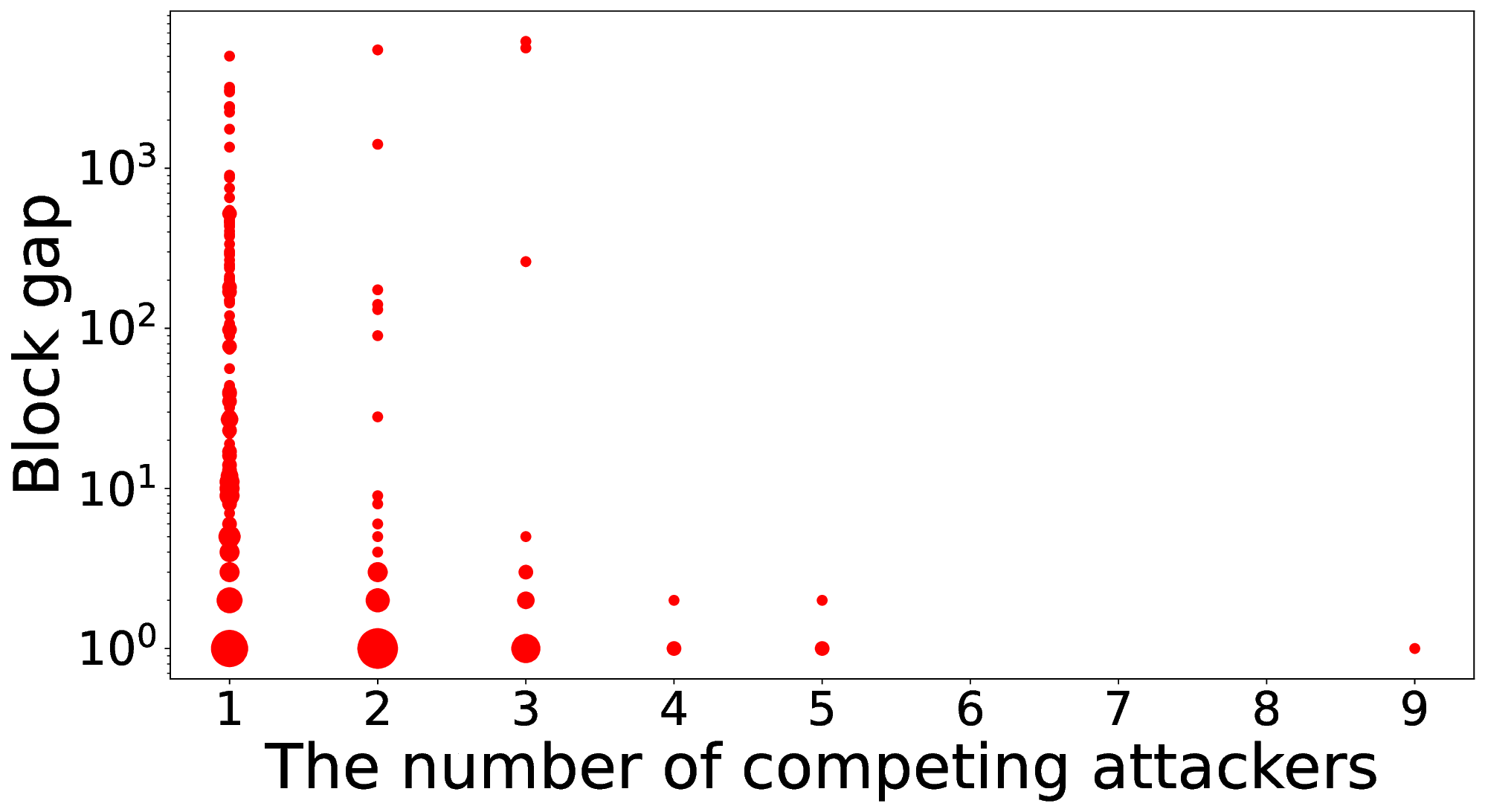}
\label{fig:blockdif_1}
    }%
    \subfloat[Block gaps and value stolen]{%
\includegraphics[width=0.235\textwidth]{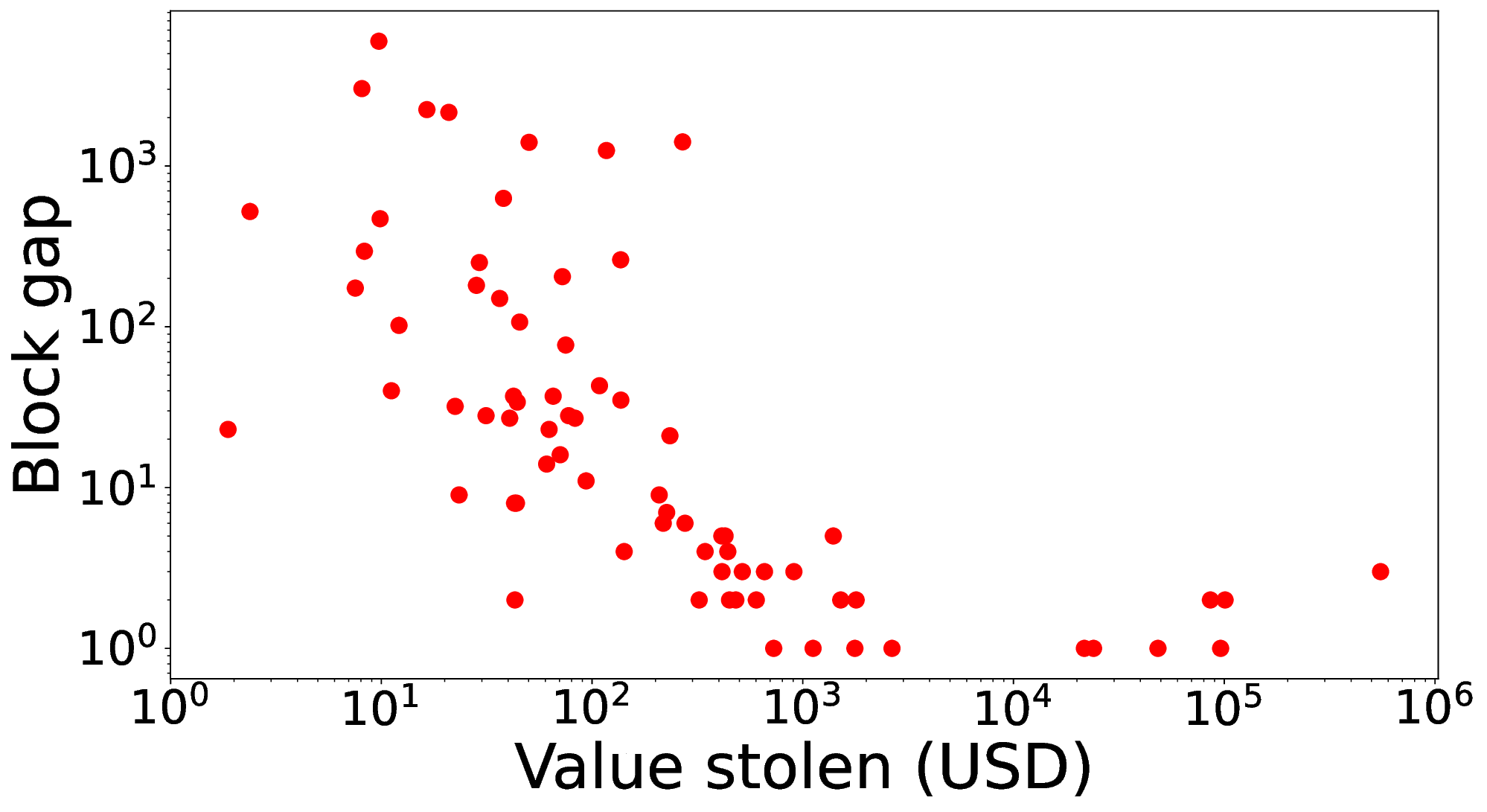}
\label{fig:blockdif_2}
    }%
  \end{center}
  \caption{Block gaps and affecting factors.}\label{fig:complex}
\end{figure}

Recall that each theft attack runs in two transactions: The first transaction that creates extractable value, and the second transaction that withdraws the value and steals it. Given each detected attack (described in the previous subsection), we measure the gap of the two blocks that include the attack's two transactions, namely block gap. We further report two metrics: the number of competing attackers and the value stolen. For each profitable deposit transaction, say $tx$, we find transactions sent by competing attackers. Here, a competing attacker either directly sends a transaction to withdraw the extractable value created by $tx$ but fails or sends a transaction to read the token balance and pool reserve (which shows the intention to withdraw the value). 

We report the block gaps with the varying number of competitors in Figure~\ref{fig:blockdif_1}. A larger dot in a X/Y coordinate in the figure represents more attacks found with the same X/Y value. The result shows that when there are many competing attackers, the block gap tends to be small. This could be explained: Because the first attacker wins, the more attackers there are, the earlier the withdrawal transaction is sent, namely a smaller block gap.
The result also shows that with few competing attackers comes a large range of block gaps.

We then report the block gaps with varying value stolen in Figure~\ref{fig:blockdif_2}. The result shows that when the stolen value is large, the block gap tends to be small. A possible explanation is that the higher extractable value gives the attacker more incentives to send the withdrawal transaction with higher fees and frontrun other thieves.
 
\subsubsection{Attacker/Victim Behavior over Time}

\begin{figure*}[!ht]
\begin{center}
\includegraphics[width=0.7\textwidth]{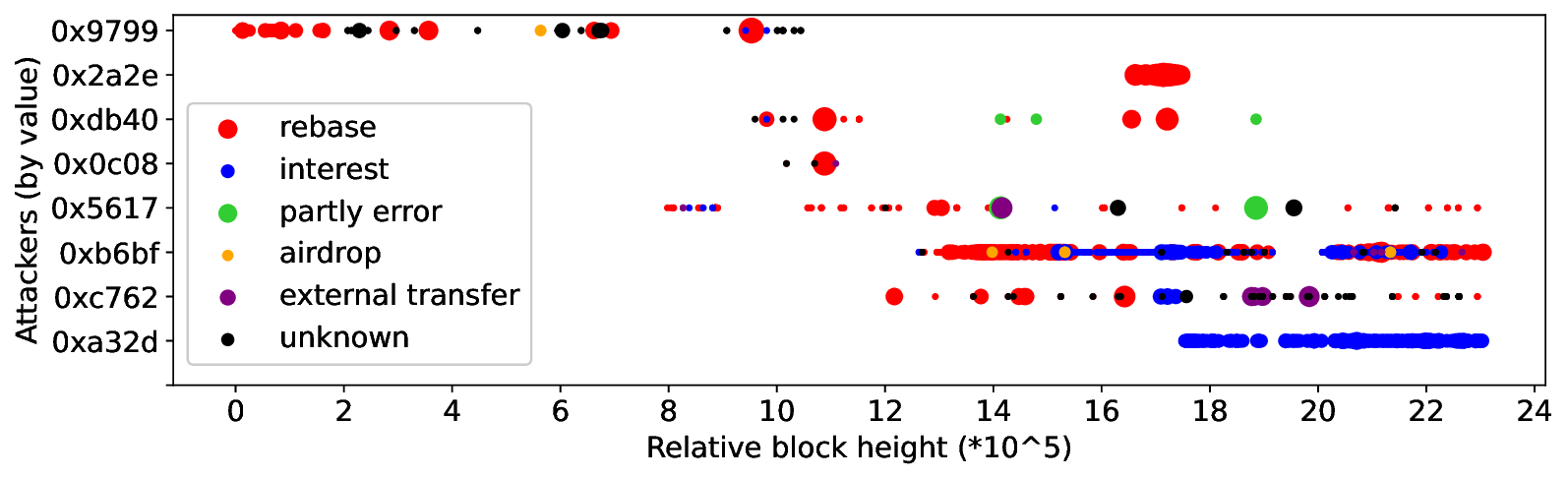}
\end{center}
\vspace{-0.1in}
\caption{Attack timeline of top-8 attackers. In the Y axis, attacker accounts are ordered by the total value stolen.
Data from 07/09/2020 (block 10422671) - 06/29/2021 (block 12727620).}
\label{fig:attack}
\end{figure*}

\begin{figure*}[!ht]
\begin{center}
\includegraphics[width=0.7\textwidth]{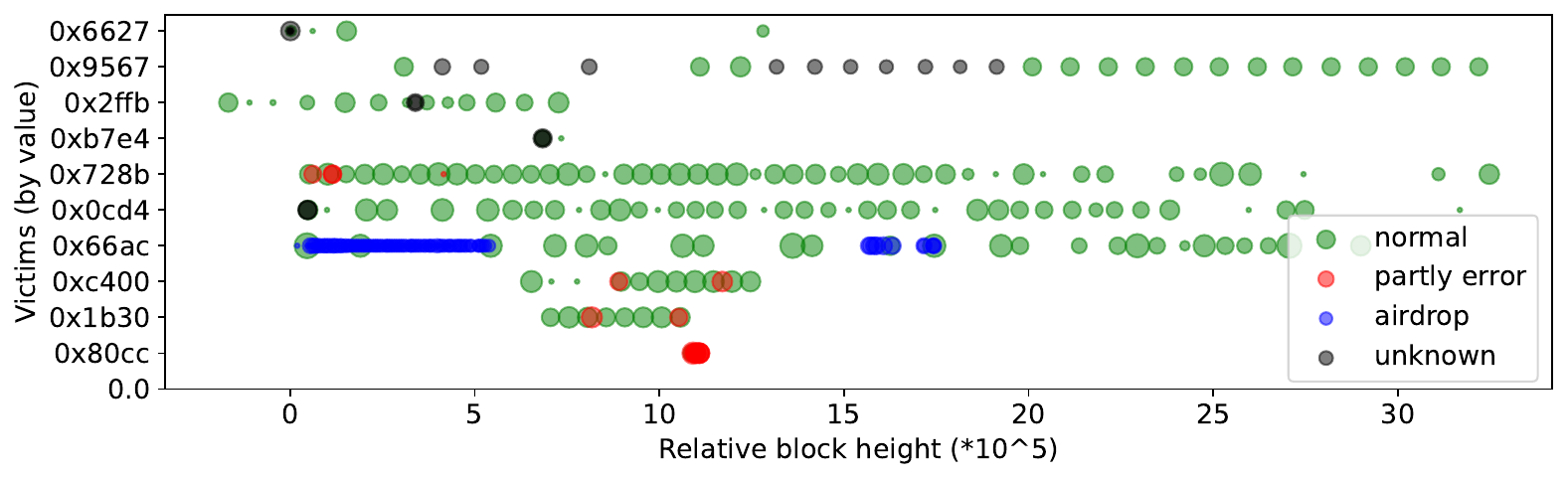}
\end{center}
\vspace{-0.1in}
\caption{Attack timeline of top-10 victims. In the Y axis, victim accounts are ordered by the total value stolen.
Data from 07/29/2020 (block 10553883) - 01/10/2022 (block 13979019).}
\label{fig:victim}
\end{figure*}

We plot the timeline of the top-8 attackers' activities in Figure~\ref{fig:attack}. The attackers are ranked by the total value stolen, with attacker $0x9799$ being the top-1. Each dot represents a detected attack sent by the attacker's account. The size of the dot represents stolen value. Color represents the exploit pattern. The most successful attacker ($0x9799$) steals value through different causes (including rebase, airdrop, etc.) The attacker's activities last about 
$1.1*10^6$ blocks, which amount to six months. The second attacker ($0x2a2e$) concentrates all her attacks on a one-month span. She exploits exclusively the value created by rebase operations. The third attacker ($0xdb40$) mounts attacks sporadically and target value created through two causes: Token interest and rebase operations. The ninth attacker exploits buggy smart contracts to gain illicit profit. The tenth attacker exclusively grabs the value from the token interest in a three-month period. As a result, most top attackers mount multiple attacks consistently, which shows the attacks are intended and not by accident.

We additionally study top victims' behavior. The victims are ranked by the value stolen. In Figure~\ref{fig:victim}, we plot in dots the transactions sent by victims whose deposits are stolen in identified attacks. Different color represents different causes. Additionally, in hollow green circles, we plot the normal transactions sent by the victims and related to Uniswap. The result shows that most victims change behavior after an attack/theft. For instance, victim accounts, including $0x0cd4$ and $0x80cc$, all stop using the vulnerable interface (e.g., partly error) to send a deposit after an attack. However, other victims can repeatedly create extractable value that gets stolen. For instance, account $0x66ac$ repeatedly airdrops tokens to the Uniswap pool, which gets stolen. 

\subsubsection{Value Received by Pools}

\begin{table}[!htbp] 
\caption{Extractable value and its destinations.}
\label{tab:extractable_value}
\centering{
\begin{tabularx}{0.425\textwidth}{ |X|l|X| } 
\hline
Value extracted by & Rebase ($P_1$) & External transfer ($P_3$) \\
\hline
External attackers & \$0.81M & \$0.50M \\
\hline
Pools (intended) & \$58.1M & \$1.22M \\
\hline
Pools (accidental) & \$0.65M & \$1.29M \\
\hline
\end{tabularx}
}
\end{table}

\begin{figure}[!htbp]
\begin{center}
  \subfloat[Value created by rebase ($P_1$)]{%
     \includegraphics[width=0.245\textwidth]{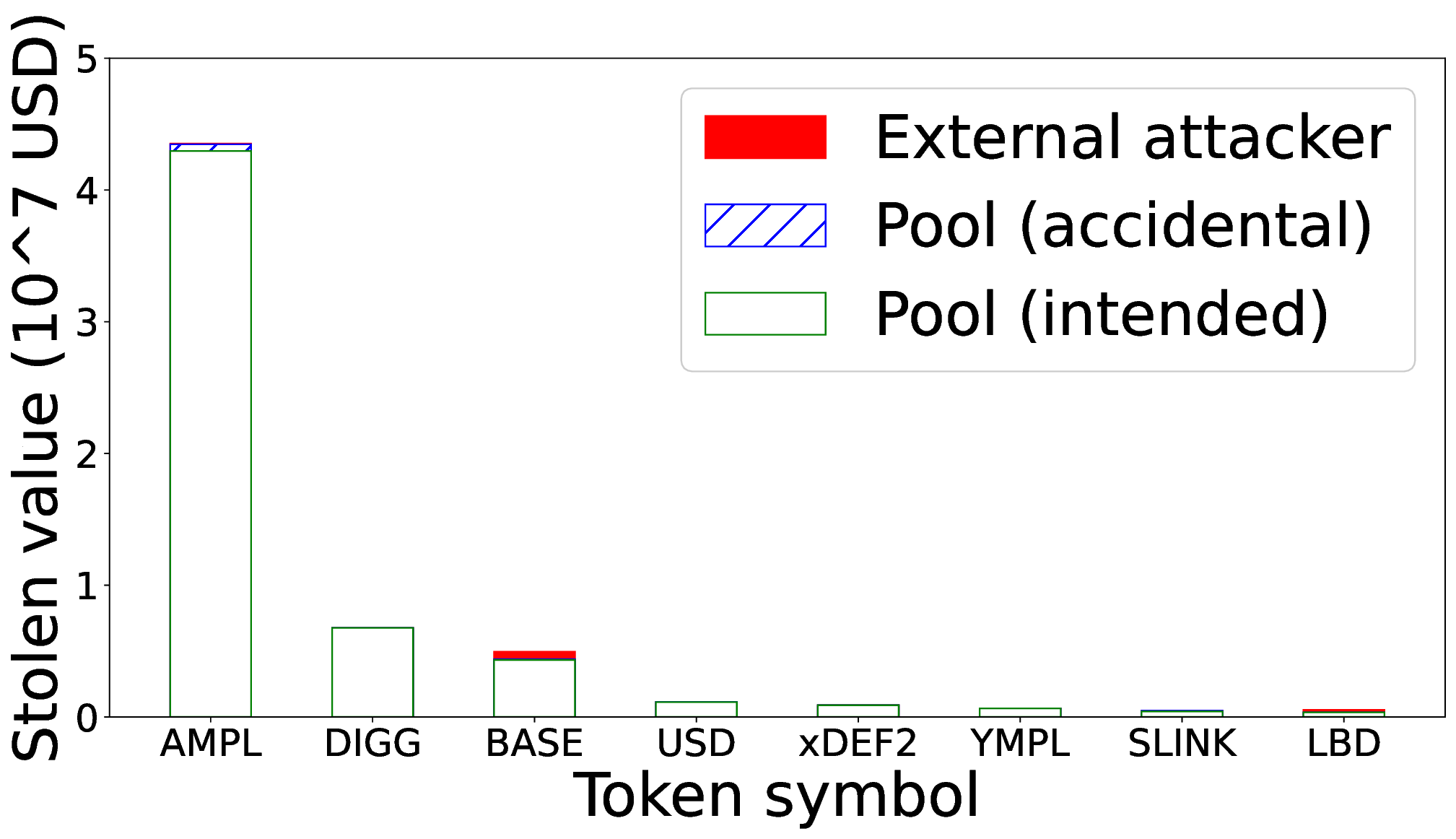}
    \label{fig:moneyflow_1}
  }%
  \subfloat[Value created by external transfers ($P_3$): all pools are between token $XX$ and $WETH$, where $XX$ is the tick on the $x$ axis.]{%
    \includegraphics[width=0.245\textwidth]{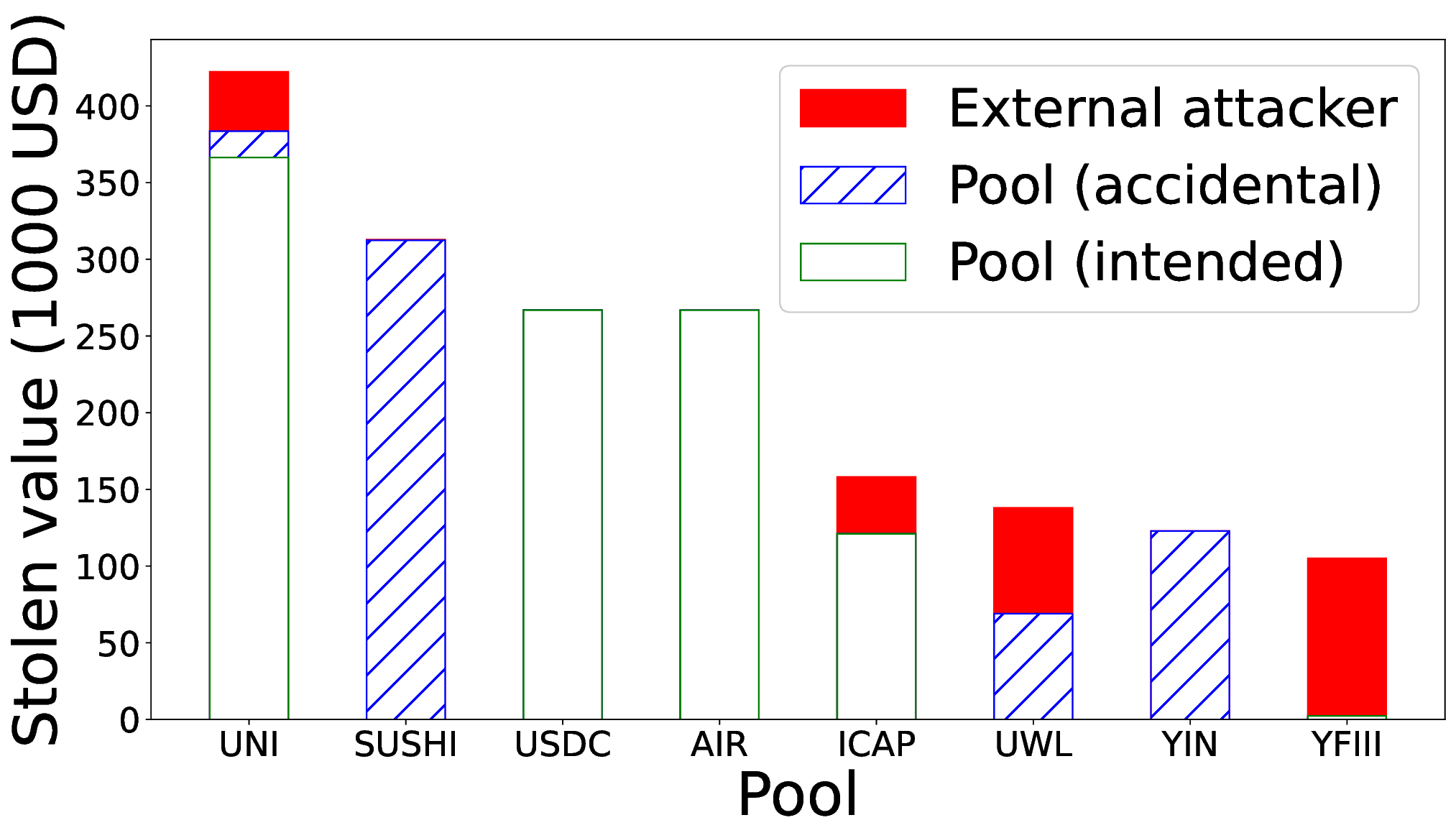}
     \label{fig:moneyflow_2}
     }%
\end{center}
\caption{Extractable value received by the pool versus that by attacker accounts.}\label{fig:complex}\vspace{-0.10in}
\end{figure}

Extractable value can be claimed by any account, including an external attacker and the pool itself. Our measurement result so far captures the former case. In this experiment, we measure the extractable value that goes to the pool.

A pool can claim extractable value by calling the \texttt{sync} function, which synchronizes the pool reserve and token balance. In Uniswap, there are three circumstances \texttt{sync} is called: a standalone call to \texttt{sync} in a transaction, \texttt{sync} called after a \texttt{swap} in the same transaction, \texttt{sync} called after an \texttt{addLiquidity} call in the same transaction. In the first case, tokens are sent to the pool intentionally (denoted by ``pool (intended)'' in the table and figures), and in the second and third cases, the pool receives deposit by accident (denoted by ``pool (accidental)''), because it is an irrelevant transaction to \texttt{sync} the reserve and token balance. Informally, we calculate the extractable value that goes to the pool in the following method: We obtain the value of the pool reserve before and after the transaction. We also parse the function call arguments from the transaction data field. We can cross-check the pre-state, post-state, and function arguments to detect the mismatch, which can be attributed to the extractable value. For instance, a normal \texttt{swap} in Uniswap does not change the product in AMM, and a spurious \texttt{swap} that the pool receives extractable value can be detected by the violation of product equality.

Using the above method, we measure the total extractable value claimed by the pool under different patterns, including $P_1$ and $P_3$. The result is presented in Table~\ref{tab:extractable_value}: The total value claimed by the pools is consistently more than $5\times$ of that by external attackers. This suggests ``room'' for the current Uniswap thief to increase their revenue. We further show the total extractable value by different pools/tokens. In Figure~\ref{fig:moneyflow_1}, we plot the extractable value created by different rebase-enabled tokens (Pattern $P_1$). 
Token $AMPL$ has the highest extractable value, most of which goes back to the pool by the token issuer intentionally calling \texttt{sync} function after each \texttt{rebase} call (the ``intended'' case). 
Token $BASE$ is the third highest in total extractable value and has the highest value stolen by external attackers. In Figure~\ref{fig:moneyflow_2}, we plot the extractable value created by external transfers (Pattern $P_3$) against different pools. Here, all the top-value pools we found are between token $WETH$ and another token. The most vulnerable pool is between Token $UNI$ and $WETH$, whose extractable value is $400$ thousand USD in total, among which the $10\%$ value is extracted by external attackers and the rest is mostly sent to the pool intentionally. Token $ YFill$'s value is exclusively extracted by external attackers. Token $ SUSHI$'s deposit is mostly claimed by the pool accidentally. 

\section{Detecting Lost Tokens}
\label{sec:losttokens}

\noindent
{\bf Problem}:
We detect lost tokens out of standalone deposits (i.e., fairness violation II). On the one hand, the incidents of lost tokens appear as standalone deposits. On the other hand, not every standalone deposit should be attributed to lost tokens. There are other causes, such as token giveaways, and administrative accounts calling privileged AMM operations. 

{%
\noindent
{\bf Methods}: 
We propose two heuristics for detecting lost tokens. 
The first detection heuristic is based on the following intuition: a trader account may use an unsupported mechanism to deposit tokens. She may then realize the mistake and retry the token deposit of the exact same amount using the right mechanism. For instance, Uniswap V3 does not support deposit by \texttt{transfer}. Unaware of the right function, a user may initially call \texttt{transfer} to deposit a token to a Uniswap V3 pool. The user then switches to calling \texttt{approve} of the same amount for the second time. The \texttt{approve} call is accepted by the V3 pool. Based on the heuristic, we propose the following indicator $I_9$:

\begin{center}\fbox{\parbox{0.90\linewidth}{
Indicator $I_9$: 
Suppose an account Alice sends a standalone deposit (Violation II) of $x$ units of Token $T_0$ to a pool. Alice then sends a fair swap of the same $x$ units of $T_0$ but using a different mechanism from the standalone deposit. The sequence indicates that the first standalone deposit results in lost $x$ units of Token $T_0$.
}}\end{center}
}

In the second heuristic, the idea is that lost tokens must be deposited by an unprivileged AMM account. And a privileged account, such as an AMM-pool administrator, is entitled to deposit value to an AMM pool, which is unlikely a mistaken deposit. We propose the indicator for lost token below:

\begin{center}\fbox{\parbox{0.90\linewidth}{
Indicator $I_10$: An account that calls privileged AMM functions (e.g., \texttt{setSwapFee()} in Balancer) is an privileged account, and a standalone deposit sent by a privileged account is not attributed to lost tokens.
}}\end{center}

\begin{table}[!htbp] 
\caption{Aggregated results on lost tokens.}
\label{tab:lost}\centering{\small
\begin{tabularx}{0.5\textwidth}{ |X|l|l|l| }
  \hline
   AMM & \# txs & Total value (USD) & Block diff (avg) \\ \hline
   Uniswap-V3 & $20$ & $47928$ & $9013$ \\ \hline
   Balancer & $192$ & $7875$ & $6578$ \\ \hline
   Curve & $3$ & $905$ & $14533$ \\ \hline
\end{tabularx}
}
\end{table}

\noindent
{\bf Results}:
Using the method above, we detect $215$ instances of lost tokens on Uniswap V3, Balancer, and Curve, inflicting a total of lost tokens worth $56708$ USD, as shown in Table~\ref{tab:lost}. Given a standalone deposit, we calculate its value using the token price at the time of the deposit (by using the nearest price with token WETH in Uniswap V2). We record the block difference between the standalone deposit and the follow-up deposit retry sent from the same account. It can be seen block difference on Uniswap V3 is about $9013$ blocks which amounts to about $33$ hours. Detailed case analysis is in Appendix~\ref{sec:losttoken:cases}.

{%
\section{Root Causes \& Countermeasures}
\label{sec:mitigate}

We first present a security analysis of different AMM protocols under the risks of token thefts and lost tokens. We then present our two mitigation designs, a secure redesign of AMM pool and a retrofittable scheme to harden deployed pools' security. 

\subsection{Risk Analysis}
\label{sec:riskanalysis}

\begin{table*}[ht]
\begin{center}
\begin{minipage}{.46\textwidth}
 \includegraphics[width=0.95\textwidth]{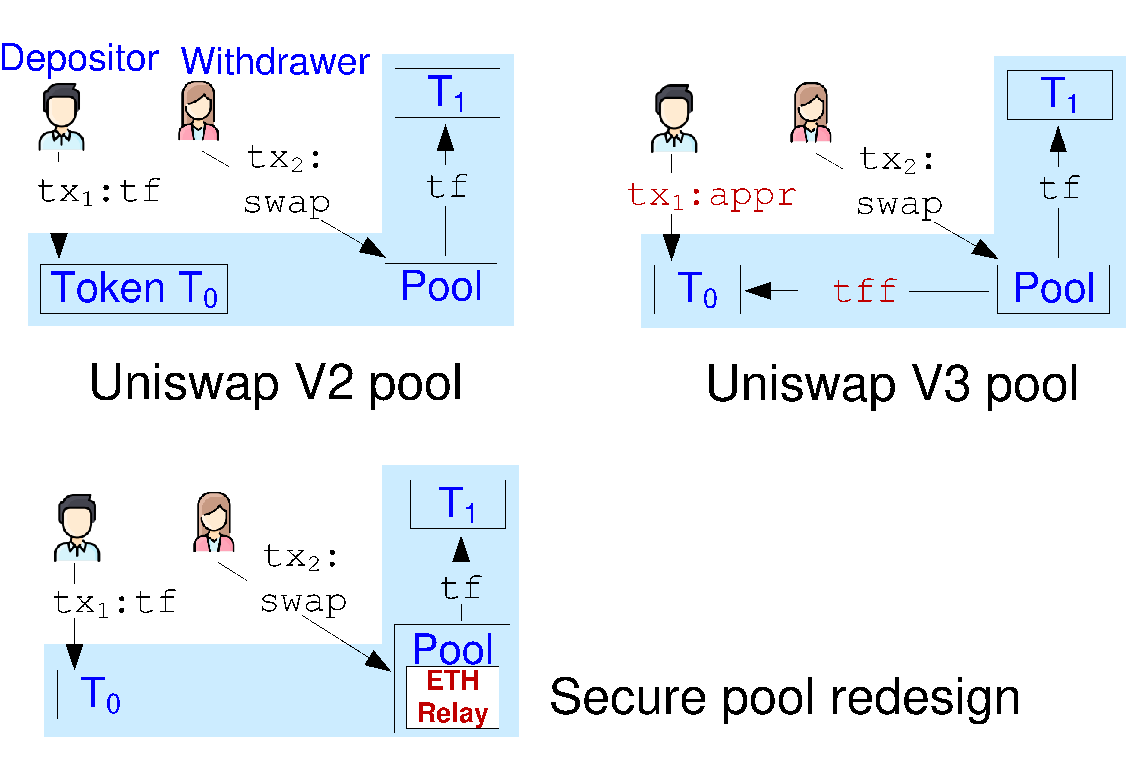}
 \captionof{figure}{AMM pool designs: \texttt{tf}/\texttt{off}/\texttt{appr} refer to ERC20 functions \texttt{transfer}/\texttt{transferFrom}/\texttt{approve}.}
  \label{fig:ammpool:designs}
\end{minipage}
\hspace{0.1in}
\begin{minipage}{.5\textwidth}
  \centering{\footnotesize
  \begin{tabular}[b]{lcc}
  \hline 
  & Direct deposit by & Indirect deposit by  \\ 
  & \texttt{transfer} & \texttt{transferFrom} \\ \hline 
  \multicolumn{3}{l}{{\it Uniswap V2}}\\ \hline
  \ \ \ Swap tokens & \cmark & \xmark  \\ \hline 
  \ \ \ Secure swap against theft & {\color{red} \xmark} & \cmark (N/A) \\ \hline 
  \ \ \ Refund against lost tokens & \cmark & \cmark  \\ \hline 
  \multicolumn{3}{l}{{\it Uniswap V3}} \\ \hline
  \ \ \ Swap tokens & \xmark & \cmark  \\ \hline 
  \ \ \ Secure swap against theft & \cmark (N/A) & \cmark \\ \hline 
  \ \ \ Refund against lost tokens & {\color{red} \xmark}  & {\xmark} \\ \hline 
  \multicolumn{3}{l}{{\it Secure AMM pool redesign (\S~\ref{sec:redesign})}}\\ \hline
  \ \ \ Swap tokens & \cmark & \cmark  \\ \hline 
  \ \ \ Secure swap against theft & \cmark & \cmark \\ \hline     
  \ \ \ Refund against lost tokens & \cmark & \cmark \\ \hline 
  \end{tabular}%
  }
\caption{Risk analysis (red cross marks are risks that pose challenge to fix).}
\label{tab:ammsecurity}
\end{minipage}
\end{center}
\vspace{-0.2in}
\end{table*}

\ignore{
\begin{figure}[!ht]
\begin{center}
\includegraphics[width=0.3\textwidth]{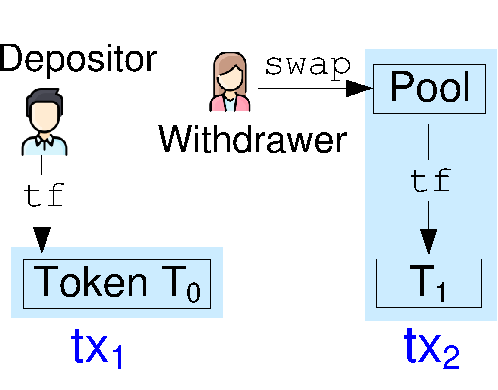}
\end{center}
\caption{The AMM service supporting token deposit-by-\texttt{transfer}; the functions are specific to the example of Uniswap V2.}
\label{fig:ammpool:designsXX}
\end{figure}
}

  Uniswap V3 that only accepts indirect deposit by \texttt{approve} and \texttt{transferFrom}, 
  and our secure pool design that supports both direct and indirect deposits with security against theft.

We character AMM protocol by how token deposits are supported. One design adopted in the Uniswap V2 family is to accept direct token deposit by \texttt{transfer}. This is depicted in Figure~\ref{fig:ammpool:designs}, where a trader deposits token $T_0$ by calling $T_0$ smart contract's \texttt{transfer} function. Another design, adopted by Uniswap V3, Balancer and Curve, is to accept {\it indirect token deposit} by \texttt{approve}/\texttt{transferFrom}. In this design, a trader first calls \texttt{approve} to delegate the spending of her token $T_0$ to the pool which then does the swap fairly by calling $T_0$'s \texttt{transferFrom} and $T_1$'s \texttt{transfer} in the same transaction. This is depicted in Figure~\ref{fig:ammpool:designs}.

In the following, we analyze the security of Uniswap V2 as a representative pool supporting direct deposits and the security of Uniswap V3 as a representative pool supporting indirect deposits. 

\noindent
{\bf Risk of Uniswap V2}: Uniswap V2 is insecure against token thefts from direct deposit. Once its pool contract account receives a token deposit via \texttt{transfer}, the deposited value can be withdrawn by any subsequent transaction (i.e., calling \texttt{swap} function). We call this vulnerability by permissionless withdrawal. As analyzed later, this vulnerability applies to any direct-deposit based AMM pool design.

Uniswap V2 supports token refund for standalone (direct) deposit via \texttt{skim} function. For indirect deposit, Uniswap V2 does not allow withdrawal and is thus secured against possible token thefts. Since indirect deposit is through \texttt{approve}, a standalone \texttt{approve} does not decrease the depositor trader's balance. Thus, Uniswap V2 can ``refund'' mis-issued deposit. Note that we scan the Uniswap V2 smart contract code to verify that it does not issue any \texttt{transferFrom} call.

\noindent
{\bf Risk of Uniswap V3}: Uniswap V3 is secure against token thefts. When the deposit is directly made via \texttt{transfer}, V3 does not support it and thus does not allow any withdrawal from it. When the deposit is indirectly made via calling \texttt{approve} function, V3 does allow withdrawal and is secured against theft. This is because with indirect deposit the actual token deposit (done in \texttt{transferFrom}; recall Figure~\ref{fig:ammpool:designs}) is executed in the same transaction with token withdrawal, enforcing both fairness (against a thief trader's attempt to insert a malicious withdrawal in between) and the match between the deposit sender and withdrawal receiver.

Uniswap V3 does not refund mistaken deposit, regardless whether made directly or indirectly. Because Uniswap V3's pool does not support traders canceling an \texttt{approve} call, as confirmed by the Uniswap team~\cite{me:bugreport:defiant}.

Overall, the risk analysis of Uniswap V2/V3 is presented in { Table~\ref{tab:ammsecurity}}. Cross marks refer to risky operations in the AMM pool. The black ones are easy to fix; for instance, canceling the allowance by a previous \texttt{approve} call is straightforward. More challenging are the two red cross marks associated with direct deposit, that is, to secure a pool supporting direct deposit and token refunding against token theft (as in Uniswap V2).

We additionally characterize other AMM protocols by their support of direct deposits (by \texttt{transfer}) in Table~\ref{tab:ammprofile}. All other AMM instances that don't support direct deposit are subject to the risk of lost tokens.

\begin{table}[!htbp] 
  \caption{High-level characterization of AMM protocols. ``Lost'' refers to lost tokens.}
  \label{tab:ammprofile}
  \centering{\footnotesize
  \begin{tabularx}{0.5\textwidth}{ |X|c|c|c|c| }
  \hline
Design & Uniswap V2 & Uniswap V3 & Balancer & Curve \\ \hline
Direct deposit by \texttt{transfer} 
& \cmark & \xmark & \xmark & \xmark \\ \hline
Risks
& Theft & Lost & Lost & Lost \\ \hline
  \end{tabularx}
}
\end{table}

{
\noindent{\bf 
Risk generalizability}: The security risks we discovered in this work stem from the design flaws of ERC20 smart contracts, including both information non-traceability of \text{transfer} in the ERC20 standard (see example below) and exhibiting behavior undefined in the standard.   Thus, the risks are generic and are applicable to any DeFi smart contracts built on ERC20 tokens. 
For instance, implementing the protocol-level fair exchanges~\cite{DBLP:conf/ccs/CampanelliGGN17,DBLP:journals/corr/abs-1911-09148,DBLP:conf/ndss/MalavoltaMSKM19,DBLP:conf/ccs/0001MM19,DBLP:journals/corr/MillerBKM17,DBLP:conf/podc/Herlihy18,DBLP:conf/ccs/CampanelliGGN17}, as surveyed in \S~\ref{sec:rw}, on widely deployed ERC20 tokens on Ethereum could face the same security risks uncovered from this work. 

We manually validate this claim by running an HTLC smart contract\footnote{\url{https://github.com/chatch/hashed-timelock-contract-ethereum}} atop ERC20 tokens with the following tests: An account that deposits value (mis)using ERC20's \texttt{transfer} function (instead of \texttt{approve}-\texttt{transferFrom}) is unable to call either the \texttt{withdraw} function or \texttt{refund} function with success. This implies the mis-deposited tokens are lost.

In general, given a direct deposit made by calling an ERC20 token's \texttt{transfer} function, the DeFi smart contract can decline it (as in Uniswap V3, which leads to lost tokens) or accept it to authorize the withdrawal of the other token. For the latter design, because the call of the token's \texttt{transfer} function does not notify context information (e.g., the \texttt{transfer} sender) to any third-party contract, the DeFi contract cannot determine who the original token depositor is and has to allow permissionless withdrawal, as in Uniswap V2.
}

{
\subsection{Countermeasure Design Space}
\label{sec:counter:space}

\noindent{\bf 
Existing MEV mitigations}: In the existing literature, protection against MEV is tackled by two design paradigms: 1) designing secure DeFi protocols to eliminate the profitable opportunities and {\it prevent} transaction reordering attacks (by making them unprofitable), and 2) detecting and {\it mitigating} the transaction reordering attacks (by enforcing fair transaction ordering). 

The approaches in the first paradigm are application specific (e.g., protecting AMM pools, such as rerouting against arbitrage~\cite{DBLP:journals/corr/abs-2106-07371}, may not be transferable to protecting lending services). In particular, flashbots and MEV redistribution~\cite{DBLP:conf/ccs/ChitraK22
} are based on the assumption of a trusted off-chain infrastructure that aggressively collects the MEV ahead of adversaries. The second paradigm is generic to DeFi applications, but the current approaches are limited to mitigating frontrunning attacks. Specifically, frontrunning attacks are mitigated by enforcing fair transaction ordering~\cite{DBLP:journals/corr/abs-2203-11520} where the order in which transactions are included into blocks is the same as that transactions are submitted~\cite{me:fairsequencing}. However, backrunning attacks are not mitigated (i.e., not following the second paradigm) but are prevented in an application-specific fashion (i.e., following the first paradigm).

\noindent{\bf 
Applicability of existing mitigations}: To mitigate unfair trades, one can start by applying the existing mitigation paradigms. Specifically, one can apply fair transaction ordering techniques to mitigate the theft attacks that exercise frontrunning (e.g., Case 0 studied in \S~\ref{sec:theftinstances}). However, this approach cannot mitigate the aggressive attacker (i.e., Case 1 in \S~\ref{sec:theftinstances}) who exercises backrunning strategies. 

Alternatively, one can customize the MEV-redistribution infrastructure to capture the extractable value caused by permissionless withdrawal. In this defense, one has to assume trust in the off-chain infrastructure. We describe the detail of the custom MEV-redistribution infrastructure in Appendix~\ref{sec:hardening:mevredistribute}. 

In the rest of this section, we focus on redesigning secure AMM pools to eliminate MEV, namely following the second paradigm. Existing approaches in the second paradigm are application specific and inapplicable to the specific case of permissionless withdrawal.
} 

\subsection{Secure AMM Pool Redesign}
\label{sec:redesign}

\noindent{\bf Design goals}: We propose building a secure AMM pool that can swap or refund tokens without losing the security against token theft. This design goal is shown in Table~\ref{tab:ammsecurity}. The difficulty comes from the following: If the ERC20 tokens are directly deposited by calling token function \texttt{transfer(address from, address to, int amount)}, the pool smart contract cannot know the trader who made the deposit (recall Figure~\ref{fig:ammpool:designs} ). Without knowing the depositor account, the pool smart contract is left unable to enforce access control on where the withdrawn/refunded tokens should go. Therefore, the key of the secure pool design is to make the pool smart-contract able to verify the depositor account.

\noindent{\bf Design rationale}:
{\bf Proposed design}: Our AMM pool redesign follows the paradigm in Uniswap V3, that is, supporting swaps by indirect token deposit via \texttt{approve}/\texttt{transferFrom} calls. When an account deposits tokens directly through \texttt{transfer} calls, the AMM pool runs ETHRelay~\cite{DBLP:conf/blockchain2/FrauenthalerSSS20,me:ethrelay,me:btcrelay,me:btcrelay:github} to validate the transaction of direct \texttt{transfer} call. 
If the validation is successful, the pool extracts the original depositor's account. It enforces access controls on the token withdrawal or refund, depending on application needs, so that the original depositor receives the token refund. To save the gas fee, one can use existing cost-optimization schemes~\cite{DBLP:conf/sigsoft/WangZLTCLC21,DBLP:conf/middleware/LiTCYXX20,10018598,DBLP:conf/ccs/WangT22}.

Specifically, ETHRelay is a smart-contract procedure that takes as input an Ethereum transaction, say $tx$, and the associated proof (incl. Merkle proof within a block and a sufficient number, say $X$, of block headers). It validates the inclusion of $tx$ in the Ethereum blockchain history. The validation is secure against forged inputs based on the hardness assumption of forging $X$ Ethereum block headers that can pass validation.
The pseudocode of pool redesign and more details are described in Appendix~\ref{appdx:detail:redesign}. 

{%
The secure AMM redesign can be easily extended to support the case that the depositor designates a third-party token receiver instead of herself. The secure AMM pool can authorize token withdrawal if a receiver account is explicitly designated and signed by the depositor.}
}

\noindent{\bf Security analysis}: Suppose benign trader Alice sends a transaction $tx_1$ to directly deposit tokens (i.e., calling \texttt{transfer}). Normally, Alice would send the second transaction $tx_2$ to either request refund or finish the swap. 
Now, Adversary Bob sends a transaction $tx_2'$ to frontrun Alice's transaction $tx_2$. If Bob wants to receive the value, he needs to forge another (non-existing) transaction $tx_1'$, where Bob is the depositor account. Bob then prepares a proof of $tx_1'$'s inclusion to pass the ETHRelay validation, which cannot succeed due to the hardness assumption.

{
In the case of network congestion, the underlying consensus may cause the delayed inclusion of transactions or even dropping the transactions. Delayed or dropped transactions do not lead to unfair trades in our countermeasure, assuming that the account having deposited tokens would retry to withdraw the value. Because of the ETHRelay that prevents any other account from withdrawing the value, the retrying of withdrawals by the original depositor account can (eventually) succeed.
}

{
\noindent{\bf 
Implementation \& evaluation}: 
We implemented the secure pool redesign by patching Uniswap V3~\cite{me:mitigate:v3} and V2~\cite{me:mitigate:v2}. In the implementation, we remove two functions in Uniswap's original pool contract to make room for our code (Uniswap's original pool contract uses up the maximal $24576$ bytes allowed by an Ethereum transaction). 

We build two testing smart contracts to evaluate the security of our patched contracts. Given a pool contract, our smart contract testing lost-tokens first makes a direct token deposit via ERC20 \texttt{transfer} and then issues a token withdrawal. The success of the token withdrawal means the tested pool does not have the risk of lost tokens. 

Given a pool contact, our smart contract testing permissionless withdrawals first makes a deposit from one account and then issues a withdrawal from another account. The withdrawal failure implies the tested pool's security against permissionless withdrawals. The testing contracts are also included in the open-sourced pool patches.

We have run the two testing smart contracts against our patched pools on Uniswap V2 and V3. The results show both patched contracts are secured against lost tokens and permissionless withdrawal.
}

\section{Responsible Disclosure}
\label{sec:disclosure}

{
We disclose the bugs to both affected AMM pools and token developers after the bug discovery. The Uniswap team (\url{https://uniswap.org/bug-bounty}) has confirmed the lost-token bug on V3 but blames users' API misusing for the theft risk on V2. We did not receive feedback from the other teams we sent our report to. Bug reports we sent are documented in the anonymized Google document~\cite{me:bugreport:defiant}. The likely cause of low response rates is that the bugs we found are distinctly caused by the interaction between the pool and token smart contracts, and neither side is willing to take full responsibility for bug fixing.
}

\section{Conclusion}
\label{sec:conclude}

This paper presents the first large-scale measurement study that uncovers the prevalence of unfair trades on popular DEX services on Ethereum and Binance Smart Chain (BSC). The study unearthed $671,400$ unfair trades on all six measured DEXes, including Uniswap, Balancer, and Curve, and attribute $55,000$ instances to token thefts that inflict more than $3.88$ million USD lost. Furthermore, the measurement study uncovers previously unknown causes of extractable value and real-world adaptive strategies to these causes. Finally, we propose countermeasures to redesign secure DEX protocols and to harden deployed services against the discovered security risks.

\section{Acknowledgment}
\label{sec:acknowledgment}

All authors but the sixth are partially supported by NSF awards CNS-2139801, CNS-1815814, DGE-2104532, and an Ethereum Foundation academic grant.

\clearpage








\bibliographystyle{IEEEtranS}
\bibliography{bkc.bib,yuzhetang.bib}

\appendices
\appendices

\section{Countermeasure by Hardening Deployed AMM}
\label{sec:hardening:mevredistribute}

We propose running a lost-and-found off-chain web service. This off-chain service monitors the on-chain state for profitability and collects extractable value ahead of actual attacks. To do so, the service takes an aggressive strategy by directly monitoring the vulnerable deposits (i.e., $P_1$, $P_3$, $P_4$) and withdrawing immediately upon discovery. For token interest $P_2$, the service monitors mempool for stealing transactions sent to the pools supporting token interests. Upon the discovery, the service sends transactions to frontrun the stealing transactions. 

In essence, this service performs the same functionality as thief attackers in discovering profitability opportunities on-chain, but it is benign in the sense that it stores the extracted value and allows victims to claim their tokens. Technically, given the value deposited by Account $V$, the service stores the value and allows $V$ or any other account with $ V$'s signature to claim the value.

\ignore{
\noindent
{\bf Fuzzing router smart contracts}: For the fairness violations caused by buggy router contracts (i.e., $P_4$), one can mitigate the problem by developing more secure router smart contracts. Here, to ensure that the router calls deposit and withdrawal in all possible code paths, we fuzz the router smart contract using a fuzzing framework Echidna~\cite{DBLP:conf/issta/GriecoSCFG20}. Specifically, we develop a test oracle that defines swap fairness and a smart contract rewriter to automatically insert the oracle to the target router smart contract. The test oracle records all the \texttt{transfer}/\texttt{transferFrom}/\texttt{swap} invocations and checks if there is a standalone invocation in the end of execution.   
}

\section{Detail of Secure Pool Redesign}
\label{appdx:detail:redesign}
\lstdefinestyle{mystyle}{
    backgroundcolor=\color{backcolour},   
    commentstyle=\color{codegreen},
    keywordstyle=\color{codepurple},
    stringstyle=\color{blue},
    basicstyle=\footnotesize\ttfamily,
    breakatwhitespace=false,
    breaklines=true,
    captionpos=b,
    keepspaces=true,
    numbers=left,
    numbersep=10pt,
    showspaces=false,
    showstringspaces=false,
    showtabs=false,
}
\lstset{style=mystyle}

\definecolor{verylightgray}{rgb}{.97,.97,.97}

\lstdefinelanguage{Solidity}{
	keywords=[1]{anonymous, assembly, assert, balance, break, call, callcode, case, catch, class, constant, continue, constructor, contract, debugger, default, delegatecall, delete, do, else, emit, event, experimental, export, external, false, finally, for, function, gas, if, implements, import, in, indexed, instanceof, interface, internal, is, length, library, log0, log1, log2, log3, log4, memory, modifier, new, payable, pragma, private, protected, public, pure, push, require, return, returns, revert, selfdestruct, send, solidity, storage, struct, suicide, super, switch, then, this, throw, transfer, true, try, typeof, using, value, view, while, with, addmod, ecrecover, keccak256, mulmod, ripemd160, sha256, sha3}, 
	keywordstyle=[1]\color{blue}\bfseries,
	keywords=[2]{address, bool, byte, bytes, bytes1, bytes2, bytes3, bytes4, bytes5, bytes6, bytes7, bytes8, bytes9, bytes10, bytes11, bytes12, bytes13, bytes14, bytes15, bytes16, bytes17, bytes18, bytes19, bytes20, bytes21, bytes22, bytes23, bytes24, bytes25, bytes26, bytes27, bytes28, bytes29, bytes30, bytes31, bytes32, enum, int, int8, int16, int24, int32, int40, int48, int56, int64, int72, int80, int88, int96, int104, int112, int120, int128, int136, int144, int152, int160, int168, int176, int184, int192, int200, int208, int216, int224, int232, int240, int248, int256, mapping, string, uint, uint8, uint16, uint24, uint32, uint40, uint48, uint56, uint64, uint72, uint80, uint88, uint96, uint104, uint112, uint120, uint128, uint136, uint144, uint152, uint160, uint168, uint176, uint184, uint192, uint200, uint208, uint216, uint224, uint232, uint240, uint248, uint256, var, void, ether, finney, szabo, wei, days, hours, minutes, seconds, weeks, years},	
	keywordstyle=[2]\color{teal}\bfseries,
	keywords=[3]{block, blockhash, coinbase, difficulty, gaslimit, number, timestamp, msg, data, gas, sender, sig, value, now, tx, gasprice, origin},	
	keywordstyle=[3]\color{violet}\bfseries,
	identifierstyle=\color{black},
	sensitive=false,
	comment=[l]{//},
	morecomment=[s]{/*}{*/},
	commentstyle=\color{gray}\ttfamily,
	stringstyle=\color{red}\ttfamily,
	morestring=[b]',
	morestring=[b]"
}

\lstset{
	language=Solidity,
	backgroundcolor=\color{verylightgray},
	extendedchars=true,
	basicstyle=\footnotesize\ttfamily,
	showstringspaces=false,
	showspaces=false,
	numbers=left,
	numberstyle=\footnotesize,
	numbersep=9pt,
	tabsize=2,
	breaklines=true,
	showtabs=false,
	captionpos=b
}

\begin{lstlisting}[language=Solidity,
    caption=Psuedocode of secure AMM protocol,
    linerange={1-30},
    firstnumber=1,
    deletekeywords={[2]INT},
    morekeywords={clustered},
    framesep=2pt,
    xleftmargin=5pt,
    framexleftmargin=5pt,
    frame=tb,
    framerule=0pt, 
    label=Secure_AMM,
    float,floatplacement=H
    ]
contract pool{
 ...
function refundMisDeposited(bytes32 tx1, bytes32[] block_headers, bytes32[] merkle_proof) {
  address sender,token;
  uint256 amount,value;
  if(length(block_headers) < 12) revert();
  if(!verifyHeaders(block_headers)) revert();
  if(!verify(tx1,block_headers.end(),merkle_proof)) revert();
  if(tx1.receiver != this & tx1.calldata.to()!=this) revert();
  // take action to refund
  user = tx1.sender;
  token = tx1.receiver;
  amount = tx1.calldata.amount();
  value = tx1.value;
  //refund ETH
  if(tx1.value != 0)
    tx1.sender.transfer(tx1.value);      
  //refund Token
  if(tx1.calldata.amount()!=0)
    tx1.receiver.transfer(user,amount);
}}
\end{lstlisting}

We propose an ETHRelay-style~\cite{me:btcrelay,me:btcrelay:github} protocol to refund mis-deposited value to the pool. As shown in Figure~\ref{fig:ammpool:designs}, suppose account Alice sends a transaction $tx_1$ that misuses unsupported API to make deposits (e.g., \texttt{token.transfer()} or transferring Ether in a transaction). After realizing her mistake, Alice sends the second transaction $tx_2$ to call a \texttt{refund} function in the pool. The \texttt{refund}'s signature is the following: \texttt{refund($tx_1$, block\_headers, merkle\_proof)}. Inside, the \texttt{refund} function verifies the given block headers, the amount of which should be sufficient for finality of transaction inclusion (e.g., $12$ blocks or more). The function also checks if $tx_1$ is included in the latest block header given using the \texttt{merkle\_proof}. The function further checks whether the transaction deposits value to the pool smart contract. If so, the function extracts the sender of $tx_1$, the type and amount of value deposited. If it is a token, the function then calls \texttt{transfer} function in the token smart contract to refund the tokens. If it is an Ether, the function calls the builtin \texttt{transfer} to refund Ether. In both cases, the refund is received by the sender extracted from $tx_1$, but the sender of $tx_2$. After that, the function records the refund to prevent duplicated refunds in the future.

In the above protocol, after Alice's $tx_1$, suppose another account Bob sends $tx_2'$ that frontruns Alice's $tx_2$. Bob can only ask to send the refund back to Alice, because it is hard to forge enough blocks that include a forged mis-deposit transaction $tx_1'$ with him as the sender.  

\section{Algorithm for Discovering Fairness Violations}
\label{sec:algorithm:detect}

\noindent{\bf The algorithm}:
The proposed algorithm in Listing~\ref{code:findmismatch} takes as input the traces of deposit and withdrawal records and produces as output the risky swaps it finds. The algorithm internally runs in two rounds. In the first round, it equi-joins the deposit records with withdrawal records based on their transaction IDs. In each joined tuple, which is a set of deposit records and withdrawal records in the same transaction, it matches the deposit and withdrawal records. In the removeMatch function, it finds a group of deposit records that match a group of withdrawal records in value. Specifically, if the deposit record has the same value with a withdrawal record in the same transaction, it constitutes a normal swap or addLiquidity operation, and the two records are removed from the two traces. If the value mismatches. it merges the two records into one, that is, the one with larger value. For instance, if a transfer of value 5 and a swap of value 8 are in the same transaction, it replaces the two records in the dataset with a merged swap record of value 8-5=3.

The second round runs a similar join-then-matchmaking process with the first round. The only difference is the join condition is temporal similarity, that is, a record withdrawal[i] joins deposit records that occur before withdrawal[i] and after withdrawal[i-1] (assuming deposit and withdrawal records are ordered in time). After the joined pair is produced, it runs the same match-making function and removes the matched cases.

After these rounds, what's left in the table are the mismatches.

\lstset{style=mystyle}
\begin{lstlisting}[caption=Pseudocode to find mismatch,
           %linerange={1-30},
           %firstnumber=1,
           deletekeywords={[2]INT},
           morekeywords={clustered},
           framesep=2pt,
           xleftmargin=5pt,
           framexleftmargin=5pt,
           frame=tb,
           framerule=0pt, 
           label=code:findmismatch,
           float,floatplacement=h
           ]
//deposits and swapOrAddL are both ordered by time
void findMismatch(Record[] deposits, Record[] withdrawals){
for(joinedpair in joinByTxId(deposits, withdrawals)){
   removeMatch(deposits, withdrawals, joinedpair.i1,joinedpair.i2, joinedpair.j1, joinedpair.j2);
   for(joinedpair in joinByTemporalSimilarity(deposits,withdrawals))
      removeMatch(deposits, withdrawals, joinedpair.i1, joinedpair.i2, joinedpair.j1, joinedpair.j2);
}}

void removeMatch(Record[] deposits, Record[] withdrawals,int i1, int i2, int j1, int j2){
  for (s in powerset(deposits[i1 ... i2])){
  for (t in powerset(withdrawals[j1 ... j2])){
    if (sumValue(s) == sumValue(t)){
      deposits.markRemoval(s);
      withdrawals.markRemoval(t);
      return;}}}}
\end{lstlisting}

\section{Detail of Discovered Fairness Violations}

Table~\ref{tab:violation:txnumbers} presents the number of transactions that map to fair operations and operaions violating fairness.

\begin{table}[!htbp] 
\caption{Measured fairness violations on top DEXes and their transaction numbers.}
\label{tab:violation:txnumbers}\centering{\scriptsize
\begin{tabularx}{0.5\textwidth}{ llllllX } 
\hline
 & Fair & I & II & III & IV & V \\ \hline
Uniswap-V2 & $53*10^6$ & $4767$ & $1375$ & $762$ & $3879$ & $2*10^5$ \\ \hline
Sushiswap & $3.7*10^6$ & $193$ & $61$ & $16$ & $57$ & $10$ \\ \hline
Pancakeswap & $3.0*10^6$ & {\bf $2.8*10^5$} & $2507$ & $364$ & $1.7*10^5$ & $157$ \\ \hline
Uniswap-V3 & $1.5*10^6$ & $0$ & $167$ & $0$ & $0$ & $0$ \\ \hline
Balancer & $2.29*10^6$ & $0$ & $6858$ & $0$ & $0$ & $0$ \\ \hline
Curve & $6.35*10^5$ & $0$ & $8$ & $0$ & $0$ & $0$ \\ \hline
\end{tabularx}
}
\end{table}

\section{Theft Detection Algorithm and Additional Indicators}
\label{appd:extraindicators}

\subsection{Detection Algorithm}

We design theft detection algorithm that encodes the indicators. The algorithm takes as input the detected fairness violations from the previous phase (\S~\ref{sec:detectviolatingswap}) and produces the strategies an attack account takes.

Specifically, recall that each fairness-violating swap consists of a profitable deposit and a (successful) withdrawal. We expand it by adding all the withdrawal attempts, no matter success or failure, on the same pool by calling its \texttt{getReserves} or \texttt{swap} functions between time $t_1$ and $t_2$ where $t_1$/$t_2$ is the time when the current/next deposit occurs. We ignore the fair swaps between $t_1$ and $t_2$. Hence for each violating swap, we obtain a list of accounts that send the withdrawal or expanded withdrawal attempt. For each account and each associated swap, the algorithm checks the indicators and reports the number of attacker labels the account is tagged with. In particular, indicator $I_1$/$I_2$ produces the average block gap over the account's swaps when the deposit is non-interest/interest. 

\lstset{style=mystyle}
\begin{lstlisting}[caption=Pseudocode to discover attacks,
                   %linerange={1-30},
                   numbers=none,
                   %firstnumber=1,
                   deletekeywords={[2]INT},
                   morekeywords={clustered},
                   framesep=2pt,
                   xleftmargin=5pt,
                   framexleftmargin=5pt,
                   frame=tb,
                   framerule=0pt, 
                   label=code:discoverattacks,
                   float,floatplacement=H
                   ]
void AttackDiscovery(FairnessViolation[] swaps, int th1, int th2){
  for(FairnessViolation s in swaps){
  Mapping blockGaps;

  for(Account w in s.getWithdrawers())
    if(isCA(w)) //I4
      if(findCall(w, s.getPool(), "getReserve").getTx().checkCall(w, s.pool().tokens(), "balanceOf()")) 
        emit(w, "Attacker");
    //I1
    Block depositTime=s.getDeposit().getBlock();
    Block withdrawTime=w.getBlock();
    blockGaps.getValue(w).add(withdrawTime-depositTime));
    //I2
    numberSwaps.getValue(w).increment();
  }

  for(Account w in blockGaps.keySet()){
    blockGap.getValue(w).divideBy(numberSwaps.getValue(w));
    if(blockGap.getValue(w) > th1) emit (w, "Attacker");
    if(numberSwaps.getValue(w) > th2) emit (w, "Attacker");
}}
\end{lstlisting}

\subsection{Additional Theft-Detection Indicators}

We list two additional indicators used for theft detection.

\begin{center}\fbox{\parbox{0.90\linewidth}{
Indicator $I_8$: 
An account whose withdrawal transactions appear an arbitrary number of blocks after the profitable deposits are likely to be a scavenger trader.
In other words, an account whose block gaps have a large deviation is likely to be a scavenger trader.
}}\end{center}

We propose a heuristic that intentional attackers would victimize a diverse set of accounts while a lucky trader may victimize a fixed type of accounts (e.g., in the same pool).

\begin{center}\fbox{\parbox{0.90\linewidth}{
Indicator $I_9$: If an account sends withdrawals in a large number of risky swaps, the account is an attacker.
}}\end{center}

\section{Establishing Ground Truth}
\label{sec:groundtruth}

In order to establish the ground truth whether an account $X$ is actually an attacker, we search on the Internet the webpages that link the account $X$ to a number of keywords including ``attack, vulnerable, scam, frontrunning, bot, arbitrager, MEV, bots''. Besides, we check the account $X$'s tags on Etherscan. 
Out of the $142$ suspected accounts from our detected thefts, we found $9$ accounts are complained about on the Internet webpages, $11$ accounts are tagged as miners on Etherscan, and $2$ accounts are tagged as MEV bots on Etherscan. The ground truth results are presented in Table~\ref{tab:gt}.

\begin{table*}[!htbp] 
\caption{Ground truth dataset on thief accounts}
\centering{\small
\begin{tabularx}{0.75\textwidth}{ |l|X|p{0.95cm}| } 
\hline
Account & Internet search & Etherscan tags \\ \hline
$0x9799b475dec92bd99bbdd943013325c36157f383$ & frontrunner@google & MEV bot  \\ \hline
$0x0c08545df4939ef46c1364b5930e840739667467$ & bot@google &   \\ \hline
$0x56178a0d5f301baf6cf3e1cd53d9863437345bf9$ & frontrunner@google, attack@twitter, bot@reddit &   \\ \hline
$0x00000000e84f2bbdfb129ed6e495c7f879f3e634$ & bot@google &   \\ \hline
$0xca850b6833ef86ef0484c6be74b06d61b39df031$ & scam@google &   \\ \hline
$0x17e8ca1b4798b97602895f63206afcd1fc90ca5f$ & attack@google &   \\ \hline
$0x7c651d7084b4ba899391d2d4d5d3d47fff823351$ & arbitrager@google &   \\ \hline
$0xf90e98f3d8dce44632e5020abf2e122e0f99dfab$ & mev@google &   \\ \hline
$0x42d0ba0223700dea8bca7983cc4bf0e000dee772$ & bots@google, bot@reddit &   \\ \hline
$0x0000000000007f150bd6f54c40a34d7c3d5e9f56$ &  & MEV bot\\ \hline
$0xEA674fdDe714fd979de3EdF0F56AA9716B898ec8$ &  & miner\\ \hline
$0x000000000025d4386f7fb58984cbe110aee3a4c4$ &  & MEV bot\\ \hline
$0xD224cA0c819e8E97ba0136B3b95ceFf503B79f53$ &  & miner\\ \hline
$0xb8aaed2e3117fa589eb05b0d7d0c8469d9e41ec5$ &  & miner\\ \hline
$0x829BD824B016326A401d083B33D092293333A830$ &  & miner\\ \hline
$0x5A0b54D5dc17e0AadC383d2db43B0a0D3E029c4c$ &  & miner\\ \hline
$0x99C85bb64564D9eF9A99621301f22C9993Cb89E3$ &  & miner\\ \hline
$0x1aD91ee08f21bE3dE0BA2ba6918E714dA6B45836$ &  & miner\\ \hline
$0x1aD91ee08f21bE3dE0BA2ba6918E714dA6B45836$ &  & miner\\ \hline
$0x3EcEf08D0e2DaD803847E052249bb4F8bFf2D5bB$ &  & miner\\ \hline
$0x8595Dd9e0438640b5E1254f9DF579aC12a86865F$ &  & miner\\ \hline
$0xB3b7874F13387D44a3398D298B075B7A3505D8d4$ &  & miner\\ \hline
\end{tabularx}
}
\label{tab:gt}
\end{table*}

\ignore{
\begin{table*}[h]
  \caption{Ground truth indicators}
  \label{tab:groundtruth_ind}
  \centering{\scriptsize
  \begin{tabular}{|p{6.3cm}|p{1.0cm}|p{1.7cm}|p{0.4cm}|p{0.5cm}|p{0.4cm}|p{0.4cm}|p{0.2cm}|p{0.4cm}|p{0.4cm}|p{0.5cm}|p{0.7cm}|p{0.5cm}|}
  \hline
  Accounts & Value & Deposit & \multicolumn{7}{c|}{Attack indicators} &\# &\# & Label \\
  \cline{4-10}
  (Mules) & $10^3$ USD & patterns &$I_1$ & $I_2$ & $I_8$ & $I_4$ & $I_5$ & $I_9$ & $I_6$ & pools & victims & \\ \hline
  $0x9799b475dec92bd99bbdd943013325c36157f383$ & 556 & 33$P_1$,4$P_2$,6$P_4$ &422 &371 & 1290& 0 & \xmark & 43 & \xmark & 4 & 0/7 & A3\\
  \hline
  $0x0c08545df4939ef46c1364b5930e840739667467$ & 101 & 3$P_1$,1$P_4$ &{\bf 1.5} & &0.7 & 0 & \xmark & 4 & \xmark& 1 & 1/1 & A2 \\
  \hline
  $0x56178a0d5f301baf6cf3e1cd53d9863437345bf9$ & 101 & 53$P_1$,4$P_2$,2$P_3$,2$P_4$ &13 &{\bf 1574} &24 & 0 & \xmark & 61 & \cmark & 3 & 2/4 & A1 \\
  \hline
  $0x00000000e84f2bbdfb129ed6e495c7f879f3e634$ & 0.5 & 9$P_1$,4$P_3$ &1.1 & &2.7 & 0 & \cmark & 13 & \xmark & 5 & 2/4 &  \\
  \hline
  $0xca850b6833ef86ef0484c6be74b06d61b39df031$ & 0.5 & 40$P_3$    &1.4 & &2.9 & 39 & \xmark & 39 & \xmark & 14 & 19/0 &  \\
  \hline
  $0x17e8ca1b4798b97602895f63206afcd1fc90ca5f$ & 0 & 1$P_1$       &3061& &0 & 0 & \cmark & 1 & \xmark & 1 & 0/1 &  \\
  \hline
  $0x7c651d7084b4ba899391d2d4d5d3d47fff823351$ & 1.8 & 2$P_1$     &68  & &66 & 0 & \xmark & 2 & \xmark & 1 & 1/0 &  \\
  \hline
  $0xf90e98f3d8dce44632e5020abf2e122e0f99dfab$ & 0 & 1$P_1$       &1402& &0 & 0 & \xmark & 1 & \xmark & 1 & 0/1 &  \\
  \hline
  $0x42d0ba0223700dea8bca7983cc4bf0e000dee772$ & 0.5 & 1$P_3$     &0   & &0 & 1 & \cmark & 1 & \xmark & 1 & 0/1 &  \\
  \hline
  $0x0000000000007f150bd6f54c40a34d7c3d5e9f56$ & 11 & 15$P_2$     &    &96 &0 & 15 & \xmark & 15 & \xmark & 1 & 1/0 & \\
  \hline
  $0xEA674fdDe714fd979de3EdF0F56AA9716B898ec8$ & 0.2 & 2$P_1$,5$P_2$ &0.5 &{\bf 686} &0.5 & 6 & \cmark & 1 & \cmark & 3 & 3/0 &  \\
  \hline
  $0x000000000025d4386f7fb58984cbe110aee3a4c4$ & 0.2 & 1$P_1$     &38  & &0 & 1 & \xmark & 1 & \xmark & 1 & 1/0 &  \\
  \hline
  $0xD224cA0c819e8E97ba0136B3b95ceFf503B79f53$ & 0.1 & 3$P_2$,1$P_3$ &32  &{\bf 617} &0 & 4 & \xmark & 4 & \cmark & 2 & 0/3 &  \\
  \hline
  $0xb8aaed2e3117fa589eb05b0d7d0c8469d9e41ec5$ & 0.1 & 1097$P_1$  &351 & &365 & 0 & \xmark & 1097 & \cmark & 2 & 2/0 &  \\
  \hline
  $0x829BD824B016326A401d083B33D092293333A830$ & 0.1 & 5$P_2$     &    &{\bf 696} &0 & 5 & \xmark & 5 & \cmark & 1 & 1/0 &  \\
  \hline
 \end{tabular}%
 }
\end{table*}
}

\section{Detailed Cases of Lost Tokens}
\label{sec:losttoken:cases}

\begin{table*}[!htbp] 
\caption{Top cases of lost tokens in value.}
\label{tab:lost:top}\centering{\scriptsize
\begin{tabularx}{1\textwidth}{ |X|l|l|l|l|l| } 
  \hline
Tx pair & Val. equality & Value (USD) & Block diff & AMM & Token \\ \hline
0x13743e0974d0a716bb7481841ad5f84c7304d4b1bcbfcaaec71fa9219e3652e3, 0x15642eede8d390ff8931f761c35ab5d0ca6171a177ebfbbda965e8dc91f89eb4 & True & 1609 & 20994 & Balancer & YAMv2 \\ \hline
0xc7c21765cc5f9ea252593d838df225d0d7954047a4f38f57e8635b54e3991760, 0xc7c21765cc5f9ea252593d838df225d0d7954047a4f38f57e8635b54e3991760 & True & 189 & 0 & Balancer & COMP  \\ \hline
0x90bf8271a62582cc545496b60492293e0341dbcf4f34f4b375a18998f308515e, 0x90bf8271a62582cc545496b60492293e0341dbcf4f34f4b375a18998f308515e & True & 158 & 0 & Balancer & COMP \\ \hline
0xe3b38ab3a132e198136babe2de859a13f655029e81393e240593424ffedfd465, 0x85eb4c8e793e12b86475f9e7fe5d468c7521d2fa3d5e45ca63c532705a180b1b & False & 157 & 35 & Uniswap V3 & SHIBA INU \\ \hline
0x2a79698eb4946e44b0d91ca09146045dcee2450459e98b2b780f8357a18885c8, 0x2a79698eb4946e44b0d91ca09146045dcee2450459e98b2b780f8357a18885c8 & True & 139 & 0 & Balancer & COMP \\ \hline
\end{tabularx}
}
\end{table*}

In Table~\ref{tab:lost:top}, we list top cases in value. For instance, in transaction $0x1374$ transfers, the sender account transfers $152$ YAMv2 tokens or $1609$ USD to a Balancer's pool. In this transaction, the deposit is made by directly calling the token's \texttt{transfer} function which Balancer does not take for deposit. Then, after 20994 Ethereum blocks, the sender, probably after realizing the mistake, sends another transaction $0x1564$ to call the Balancer pool's \texttt{joinpool} function to retry the deposit. The retried deposit is of the exact same amount of the same token with the original deposit. 

For another instance, in transaction $0xe3b3$, the sender transfers $4.7M$ SHIB tokens by directly calling \texttt{transfer} function to a Uniswap V3 pool which only takes deposit by \texttt{transferFrom}. After $35$ blocks, the sender switches the correct mechanism for deposit by calling Uniswap V3's \texttt{swap} function. The value in the retried deposit is twice the value in the original deposit.

\section{Extended Related Work}
\label{appdx:rw}

This work is related to blockchain data analysis. We survey blockchain-data analysis for attack detection and user de-anonymization. 

{\bf Detecting Smart-Contract Program-level Attacks}: Attack detection on blockchain generally works as distilling the attack signatures from known vulnerabilities and attack patterns and matching certain representation of smart-contract execution (e.g., in transactions, function-call graph or instruction traces) against the attack signatures. There are known patterns of smart-contract program-level attacks, such as reentrancy attacks, transaction order dependency, etc. Existing works tackle the detection of program-level attacks on Ethereum-like blockchains. 
 
TxSpector~\cite{DBLP:conf/uss/ZhangZZL20} records and replays Ethereum transactions into fine-grained EVM instruction traces, which are stored in datalog databases and queries to compare against the signatures of known attacks. Similarly, Zhou et al~\cite{DBLP:conf/uss/ZhouYXCY020} model the coarse-grained traces of smart contract function calls and match them against known attack signatures. At its core, two function calls are modeled, that is, smart contract creation/destruction and token transfers (money flow). Su et al~\cite{DBLP:conf/uss/SuSDL0XL21} uses reported attack instances to extract attack signatures, and it discovers new attacks as well as uncovers the attacker/victim behavior by systematically analyzing the Ethereum blockchain traces. Other similar work~\cite{DBLP:conf/uss/0001L21} detects the attack instances of known vulnerabilities.  

{\bf Detecting DeFi Protocol-level Attacks}: In addition to detecting program-level attacks, there are related works on detecting protocol-level attacks on deployed DeFi applications. Notable DeFi attack patterns include frontrunning, arbitrage, etc. 

Torres, et al.~\cite{DBLP:conf/uss/TorresCS21} detects front-running attacks by matching Ethereum transactions against a signature of success-failure transaction pair. Qin et al~\cite{DBLP:journals/corr/abs-2101-05511} systematizes different types of front-running attacks and measures the attacks and mitigation in the wild.

Front-running attacks are a class of attacks in which the attacker inserts a malicious transaction before the victim transaction to extract illicit profit (e.g., to win an auction bid unfairly). More generally, front-running is a class of MEV or miner extractable value, in which transaction order can be manipulated by a profit-driven miner to receive additional profit.

Arbitrage attacks work by an aggressive trader gain profit by exploiting difference of exchange rates across different DEXes/price oracles. Wang, et al~\cite{DBLP:journals/corr/abs-2105-02784} detects cyclic arbitrage on Uniswap-V2, by finding profitable cycles on token-transfer graph (nodes are tokens and edges are transfers). Similarly, Zhou, et al~\cite{DBLP:conf/sp/ZhouQCLG21} discovers profit-generating transitions in real-time by finding cycles  and leveraging solvers.

{\bf User de-anonymization}: While blockchain accounts and their activities are visible, the linkage of accounts (which are public keys) with real-world identities is not directly observable. There are existing works to infer blockchain account linkage in efforts to deanonymize users. Specifically, there are three types of linkage information: 1) Whether a specific account is linked to a specific real-world user identity. Existing works~\cite{DBLP:conf/fc/KoshyKM14} monitor a blockchain network's traffic of transactions propagated from an IP and link the transaction sender's account with the IP. 
2) Whether two blockchain accounts are linked, as the sender and receiver, in a privacy-enhanced transaction. Mixing services are intended to break the sender-receiver linkage. There are existing works aimed at reconstructing the sender-receiver relation on transactions relayed through mixing services. In particular, address reuse is a heuristic to link transaction senders/receivers on Bitcoin-like blockchains~\cite{DBLP:conf/uss/YousafKM19,DBLP:conf/uss/KalodnerMLGPCN20}.

3) Whether two blockchain accounts are shared by the same real-world user. On Bitcoin, transaction graph analysis~\cite{DBLP:conf/socialcom/ReidH11,DBLP:conf/imc/MeiklejohnPJLMVS13,DBLP:books/daglib/0040621} is proposed to recover the account linkage. Particularly, address clustering heuristics are proposed. For instance, two accounts that are co-spent by a Bitcoin transaction or that involve a change are likely to be controlled by the same off-chain user. These address clustering heuristics are useful only in the context of Bitcoin's UTXO model. Address clustering is useful to uncover the money flow in online criminal activities such as ransomware payments on Bitcoin~\cite{DBLP:conf/sp/HuangALIBMLLSM18}. 

Of most relevance to this work is the address clustering techniques proposed to link Ethereum accounts~\cite{DBLP:conf/fc/Victor20}. Specifically, the observation is that some DEXes support the deposit address, that is, trader Alice swapping token $T_0$ for token $T_1$ can specify a ``deposit'' address to receive $T_1$. Multiple accounts using the same deposit address are likely to be the same off-chain user. 

This work is different from Ethereum account linkage. In this work, the attack detection problem is to verify whether two Ethereum accounts are from two off-chain users, which is different from detecting if two accounts are linked by the same off-chain user. Moreover, the Ethereum address clustering work~\cite{DBLP:conf/fc/Victor20} has to assume that the swap is fair, normal one, so that they can attribute address sharing to the cause of the same off-chain user. By contrast, in this work, we do observe two sender accounts share the same deposit address, but they are caused by two victim accounts whose swaps are grabbed by the same attack account, as verified by our analysis of attacker/victim behavior.

 %


\clearpage

\end{document}